\documentclass[11pt]{article}
\usepackage{graphicx,cite,amsmath}

\renewcommand{\appendix}
        {
        \par
        \setcounter{section}{0}
        \setcounter{subsection}{0}
        \gdef\afterthesectionpunctdefault{:}
        \gdef\thesection{{Appendix \Alph{section}}}
        \renewcommand{\theequation}{\Alph{section}\arabic{equation}}
        \setcounter{equation}{0}
        }
\def\lsim{\hbox{\lower .8ex\hbox{$\, \buildrel < \over \sim\,$}}}
\def\gsim{\hbox{\lower .8ex\hbox{$\, \buildrel > \over \sim\,$}}}

\def\rp{\vec r_\perp}

\def\sss{\scriptscriptstyle}

\newfont{\ensmathquatorze}{msbm10 scaled 1400}
\newfont{\ensmathonze}{msbm10 scaled 1100}
\newfont{\ensmathdix}{msbm10}
\newfont{\ensmathneuf}{msbm10 scaled 833}
\newfont{\ensmathhuit}{msbm10 scaled 694}
\newfam\ensmathfam                        
\textfont\ensmathfam=\ensmathonze        
\scriptfont\ensmathfam=\ensmathdix       
\scriptscriptfont\ensmathfam=\ensmathhuit

\begin{document}

\begin{center}

{\huge Bose-Einstein beams:}\\

\

{\huge Coherent propagation through a guide}\\

\vspace{1.0cm}

{\Large P. Leboeuf and N. Pavloff}
\end{center}


\begin{center}
\noindent Laboratoire de Physique Th\'eorique
et Mod\`eles Statistiques\footnotemark,\\
\vspace{0.1 cm}
Universit\'e Paris Sud, b\^at. 100, F-91405 Orsay Cedex, France 

\vspace{2 cm}

{\bf Abstract}
\end{center}
We compute the stationary profiles of a coherent beam of Bose-Einstein
condensed atoms propagating through a guide. Special emphasis is put on the
effect of a disturbing obstacle present in the trajectory of the beam. The
obstacle considered (such as a bend in the guide, or a laser field
perpendicular to the beam) results in a repulsive or an attractive potential
acting on the condensate. Different behaviors are observed when the beam
velocity (with respect to the speed of sound), the size of the obstacle
(relative to the healing length) and the intensity and sign of the potential
are varied. The existence of bound states of the condensate is also
considered.
\vspace{3cm}
\begin{math}
\footnotetext[1]{Unit\'e Mixte de Recherche de l'Universit\'e Paris XI et du
CNRS (UMR 8626).}
\end{math}

\noindent PACS numbers:
\vspace{0.5 cm}

\noindent 03.75.b Matter waves\hfill\break
\noindent 03.75.Fi Phase coherent atomic ensembles; quantum condensation
phenomena\hfill\break
\noindent 42.65.Tg Optical solitons; nonlinear guided waves\hfill\break
\noindent 67.40.Hf Hydrodynamics in specific geometries, flow
in narrow channels\hfill\break
\newpage

\section{Introduction}

The field of Bose-Einstein condensation (BEC) of atomic vapors is undergoing
rapid experimental development providing a rich new phenomenology and also
allowing to test concepts that originated in other fields (mainly in the
theory of superfluidity, in nonlinear optics and in atomic physics).

Along these lines, the possibility of building atom lasers by guiding
condensed particles through various geometries opens up the prospects of a
rich variety of interference, transport and/or coherence phenomena. Cold atoms
have already been propagated in various guides (see e.g.
\cite{Den99,Dekk00,Key00,Mull00} and references therein); more and more
efficient coherent sources of atoms have recently been designed (using various
output coupling schemes, see Refs.~\cite{Mew97,And98,Hag99,Blo99,Blo00}) and
continuous guided beams of condensed atoms will be accessible in the near
future (see the preparatory study \cite{Man00}).

The purpose of the present work is to explicitly determine the different
propagating modes of a beam of condensed atoms through a guide, as a function
of the various external control parameters. We consider the simplest geometry,
in which a guide connects two reservoirs, and treat the case of atoms with a
repulsive effective interaction, such as rubidium and sodium. First, the
transmission through a straight guide is reevaluated: small amplitude density
oscillations, cnoidal waves, and gray solitons are possible propagating modes.
The main part of the paper is devoted to the study of coherent
transmission modes in the presence of an obstacle. We find that, as a function
of the speed of the incoming beam and the size and depth of the
perturbing potential, many different transmission modes exist. For example,
some are soliton-like modes (with a peak or a trough in the density) that are
pinned to the obstacle. They may also have density oscillations in the region
of the obstacle. On the contrary, other modes are step-like shaped. In
general, the modes may or may not have a wake. The wake, however, always
precedes the obstacle (it never occurs down-stream).

The obstacle is represented in our treatment by a potential that acts on the
condensate along the axis of the guide. It can be repulsive or attractive.
There are at least two simple ways to realize such an obstacle experimentally.
The first is to bend the guide: a potential proportional to the square of
the local curvature is created. This potential is always attractive. In view
of future applications of the atom laser to more complicated geometries, the
presence of bends seems unavoidable and their study is therefore far from
academic. The second possibility is to illuminate the beam with a detuned
laser field. Contrary to the case of a bend, attractive and repulsive
potentials may be realized by varying the frequency of the laser. This latter
method also has the advantage of allowing a better control of the relative
speed between the obstacle and the beam (if using an acousto-optic deflector).

In the case of attractive potentials, aside from the transmission modes there
may also exist bound states in the guide, in which condensed atoms are trapped
without possibility of escape. These bound states are also analyzed here. In
the case of a bend we show that, due to the intra-atomic interactions, the
bound state can support only a limited number of condensed atoms, which is
typically of the order of 100 for rubidium and sodium. However this
number can be made much larger for an attractive potential originating from a
red detuned laser field.

The paper is organized as follows. In Sec.~2 we set up the theoretical
framework and notations. Our approach for describing the condensate motion is
based on a one-dimensional reduction of the Gross-Pitaevskii equation
\cite{Jac98}. The solutions described here can in some instances be explicitly
written in an analytic form, for example, in terms of elliptic functions.
Although we use this opportunity in some cases, we have chosen to keep the
discussion at a qualitative level. This allows us to cover a large range of
experimental situations and gives a global view of all the possible solutions
in the different regimes of the control parameters. In Section~\ref{bound} we
study the existence of a bound state of the condensate created by an
attractive potential. We determine the maximum number of atoms the bound state
can accommodate. In Secs.~\ref{main}--\ref{simple} we study the propagation of
a condensate through a guide connecting two reservoirs. The analysis relies on
an interpretation of the Gross-Pitaevskii equation in terms of a fictitious
classical dynamics. We focus on different specific examples, starting with a
straight wave guide without any potential (Sec.~\ref{main}). We then study the
motion in the presence of an obstacle represented by an attractive
(Sec.~\ref{asw}) and a repulsive (Sec.~\ref{rsw}) square well.
Sec.~\ref{simple} presents two alternative treatments for determining the
transmission modes in the presence of an obstacle. The first one is a
perturbative approach; the second approximates the external potential by a
$\delta$ function. Sec.~\ref{experiments} contains an analysis of the results
in view of future experimental realizations. We present our conclusions in
Sec.~\ref{conclusion}. Some technical aspects concerning the adiabatic
approximation used for deriving the 1D reduction of the Gross-Pitaevskii
equation are included in the Appendix.

\section{An effective one-dimensional equation}\label{sec2}

We consider BEC atoms at zero temperature confined to a wave-guide. Let $x$ be
the coordinate along the axis of the guide (which is possibly bent) and $\rp$
a perpendicular vector giving the transverse coordinates. We work in the
adiabatic regime where the local curvature $\kappa(x)$ of the longitudinal
motion of the atoms is small. If the transverse extension of the wavefunction
is denoted by $R_\perp$, this is more precisely defined by the limit $R_\perp
\kappa \ll 1$ and $|d\kappa/dx|\ll 1$ (see \ref{goldstone}). In this regime
one can consider $x$ and $\rp$ as Cartesian coordinates, i.e. for instance the
volume element $d^3r$ is approximately $dx\, d^2r_\perp$.

It is consistent with the adiabatic approximation to make an ansatz for the
condensate wave-function $\Psi(\vec{r},t)$ of the form \cite{Jac98}
\begin{equation}\label{adia1}
\Psi(\vec{r},t) = \psi(x,t) \, \phi(\rp; n) \, \end{equation}
\noindent where $\phi$ is the equilibrium wave function for the transverse
motion, normalized to u\-ni\-ty, $\int d^2r_\perp |\phi|^2=1$. $\psi(x,t)$
describes the longitudinal motion, and the density per unit of longitudinal
length is
$$
n(x,t)=\int d^2r_\perp |\Psi|^2=|\psi(x,t)|^2 \ .
$$ 
Notice that the transverse wave function depends parametrically on $n(x,t)$.
The adiabatic approximation is in fact a local density approximation in the
sense that one assumes that the transverse motion is not affected by densities
at points other than $x$. As noted in Ref.~\cite{Jac98} this corresponds to
the assumption that the transverse scale of variation of the profile is much
smaller than the longitudinal one (and corresponds indeed to the limit
$R_\perp \kappa \ll 1$).

The beam is confined by a transverse potential $V_\perp(\rp)$. Keeping in mind
experimental realizations, we often consider below the particular case of a
harmonic trapping $V_\perp (\rp)= \frac{1}{2}\omega^2_\perp\, r_\perp^2$
($\omega_\perp$ is the pulsation of the harmonic oscillator; we set units such
that $\hbar=m=1$). As shown in \ref{goldstone}, within the adiabatic
approximation the presence of a bend results in an attractive longitudinal
potential $V_\parallel(x)$ given by
\begin{equation}\label{adia1a}
V_\parallel(x)=-\kappa^2(x)/8 \ .
\end{equation}

The Gross-Pitaevskii equations for the condensate are derived through a
variational principle. One extremizes the action
\begin{equation}\label{adia2}
{\cal S} = \frac{i}{2}\int d^3r\,dt 
\left(\Psi^*\partial_t\Psi-\Psi\partial_t\Psi^*\right) 
- \int dt \, {\cal E}[\Psi]
\; , \end{equation}
\noindent where ${\cal E}[\Psi]$ is the energy functional
\begin{equation}\label{adia3}
{\cal E}[\Psi] =\int d^3r \left( \frac{1}{2}|\vec\nabla \Psi|^2 + 2\pi a_{sc}
|\Psi|^4 + (V_\parallel+V_\perp)|\Psi|^2 \right) \; .
\end{equation} 
$a_{sc}$ is the s-wave scattering length of the inter-atomic potential (which
is represented by a $\delta$ function interaction). In all the present work, we
consider the case of repulsive inter-atomic interactions, $a_{sc}>0$. The
extremization of ${\cal S}$ with Lagrange multiplier $\epsilon(n)$ imposing
the normalization $\int d^2r_\perp|\Psi|^2=n$ for each $x$ (more precisely for
each $n$) leads to the following equations \cite{Jac98}
\begin{equation}\label{adia4}
-\frac{1}{2}\vec{\nabla}_\perp^2\phi +
 (V_\perp + 4\pi a_{sc} n \,|\phi|^2)\,\phi =
\epsilon(n) \, \phi
\; ,\end{equation}
\noindent and
\begin{equation}\label{adia5}
-\frac{1}{2}\partial^2_{xx}\psi + (V_\parallel + \epsilon(n))\,\psi =
 i\,\partial_t\psi \; .
\end{equation} 
In (\ref{adia5}) the non-linear term $\epsilon(n)$ (remember that
$n=|\psi|^2$) is determined as a function of $n$ from Eq.~(\ref{adia4}). In
the low density limit $a_{sc}n\ll 1$ the non-linearity in (\ref{adia4}) is
small. In this case, a perturbative solution of (\ref{adia4}) leads to
\begin{equation}\label{adia6a}
\epsilon(n)=\epsilon_{0}+2\,a_{sc} n/a_\perp^2 \ ,
\end{equation}
where $\epsilon_{0}$ is the eigen-energy of the ground-state $\phi_0$ of the
transverse unperturbed Hamiltonian $-\frac{1}{2}\vec{\nabla}_\perp^2 +
V_\perp$, and $a_\perp^{-2}=2\,\pi\int|\phi_0|^4 d^2r_\perp$. For a harmonic
confining potential $a_\perp=\omega_\perp^{-1/2}$ is known as the oscillator
length.

In the opposite large-density limit $n\,a_{sc}\gg 1$ the Thomas-Fermi
approximation holds, namely the kinetic term in (\ref{adia4}) can be neglected.
$\epsilon$ is obtained as a function of $n$ through the relation
$N_{TF}(\epsilon)=2\,a_{sc}n$, where $N_{TF}$ is the integrated Thomas-Fermi
density of states
$N_{TF}(\epsilon)=\int(\epsilon-V_\perp)\Theta(\epsilon-V_\perp)\,
d^2r_\perp/(2\pi)$. For a harmonic confining potential this reads
\begin{equation}\label{adia6b}
\epsilon(n)=2\, \omega_\perp\sqrt{n\,a_{sc}} + \epsilon_0 \ .
\end{equation}

	We remark here that the Gross-Pitaevskii equation is valid in the dilute gas
approximation, when the 3D density $n_{3\sss D}$ of the gas satisfies $n_{3\sss
D}\, a_{sc}^3 \ll 1$ \cite{Dal99}. This reads here $n\,a_{sc} \ll
(a_\perp/a_{sc})^{2/\nu}$ ($\nu=1$ in the dilute regime and $\nu=1/2$ for high
densities). $a_\perp/a_{sc}$ being typically of order $10^3$, this condition
will be considered as always fulfilled, even at high longitudinal densities,
when $n\,a_{sc}\gg 1$.

	On the other hand, the weakly interacting 1D Bose gas picture also breaks
down at very low densities, in the Tonks gas regime (recent references relevant
to this discussion are \cite{Ols98,Thy99,Pet00,Dun01}). This occurs in the
regime $n\,a_{sc}\ll (a_{sc}/a_\perp)^2 \sim 10^{-6}$ which we thus discard
from the present study.

Eq.~(\ref{adia5}), together with Eqs.~(\ref{adia6a}) and (\ref{adia6b}), are
the main results of this section. They provide an effective one-dimensional
equation for the description of the dynamics of the condensate along the
guide. Notice that in general the non-linear term in Eq.~(\ref{adia5}) does
not have the standard cubic form of the one-dimensional Gross-Pitaevskii
equation. This happens only in the low density regime $a_{sc}n\ll 1$ when the
non-linear potential (\ref{adia6a}) is proportional to $|\psi|^2$.

We emphasize that Eq.~(\ref{adia5}) relies on an adiabatic approximation. It
is well known that in many instances this approximation gives accurate results
well beyond its strict domain of validity. Two examples relevant in the
present context are Refs. \cite{Spr92} and \cite{Kas99}, where the propagation
of waves (without non-linear effects) was studied in the extreme non
adiabatic case of wave guides with a discontinuous curvature and a sudden
constriction, respectively. The adiabatic approximation was nevertheless shown
to be applicable in these systems. On the basis of these examples (and of
others), one can consider that the results presented here have a wide range of
validity.

For practical purposes it will appear useful in the following to introduce a
longitudinal healing length $\xi$ defined for a constant longitudinal
density $n$ as
\begin{equation}\label{adia7}
\frac{1}{2\, \xi^2} = \epsilon(n)-\epsilon_{0} \; .
\end{equation}
This gives $\xi=\frac{1}{2} a_\perp (n\,a_{sc})^{-1/2}$ in the low density
regime and $\xi=\frac{1}{2} a_\perp (n\,a_{sc})^{-1/4}$ for high densities in
a harmonic confining potential.

We now study in detail the different solutions of Eq.~(\ref{adia5}). Although
the attractive potential $V_\parallel(x)$ appearing in the equation of motion
was due to the presence of a bend in the guide, our results are very general
and $V_\parallel(x)$ could be of a completely different physical origin. In
particular, we will consider in Sec.~\ref{rsw} the case of a {\it repulsive}
potential which cannot be produced by a bend.

\section{Bound states}\label{bound}

We first study the existence of bound states in the guide due to an attractive
potential $V_{\parallel}(x)$. The existence of bound states in the quantum
mechanical motion of non-interacting particles in a bent wave guide has been
extensively considered in the past (see e.g. \cite{Gol92,Wu92,Lon99} and Refs.
therein). It has been shown by Goldstone and Jaffe that at least one bound
state exists in two and three dimensional bent tubes \cite{Gol92} (cf. the
discussion in \ref{goldstone}). The particle is trapped because its energy is
lower than the first propagating mode of the straight guide. In the case of a
condensed beam, we are interested in whether bound states occur in the
presence of interactions.

In the extreme dilute limit $a_{sc}n\rightarrow 0$ Eq.~(\ref{adia5}) reduces
to an ordinary one dimensional Schr\"odinger equation and the existence of a
bound state in $V_{\parallel}(x)$ is guaranteed by general theorems of quantum
mechanics \cite{Lan}. Hence there exists a state of the condensate whose
energy is lower than the energy $\epsilon_{0}$ of the first propagating mode
of the guide. This state is localized in the region where $V_\parallel$ is
noticeable, and we assume for simplicity that this happens in some finite
region around $x=0$. With increasing number of atoms in the condensate the
non-linear effects come into play. The repulsive intra-atomic interaction
increases the energy of the bound state (as well as its spatial extension),
and there is a threshold beyond which this state disappears. Therefore for a
sufficiently large number of atoms no bound-state is expected to occur. We now
determine the threshold quantitatively by determining the maximum number of
Bose condensed atoms the bound-state can accommodate.

Near the threshold the state is very weakly bound, and the wave function
extends over distances much greater than the range of the potential
$V_\parallel$. Hence, in this limit it is legitimate to make the approximation
$V_\parallel(x)\approx \lambda\,\delta(x)$, with $\lambda =
\int_{-\infty}^{+\infty} V_\parallel(x) \ d x<0$. This approximation is not
contradictory with the assumption of adiabaticity of the motion.

We look at stationary solutions of Eq.~(\ref{adia5}) of the form $\psi
(x,t)=A(x)\exp(-i\,\mu\, t)$, $\mu$ being the chemical potential and $A$ a
real function. In the regions where the potential is negligible (i.e., for
$x\neq 0$ with the above replacement of $V_\parallel$ by a $\delta$ function)
Eq.~(\ref{adia5}) can be integrated once, giving an equation for $n(x)=A^2$,
\begin{equation}\label{bound1}
-\frac{n^{\prime\,2}}{8\, n} + \varepsilon(n) = \mu \, n \; ,
 \qquad\mbox{where}\qquad 
\varepsilon(n) =\int_0^n\epsilon(\rho)\, d\rho \; .
\end{equation}
With the convention defined in Section 2, the normalization is
$\int_{-\infty}^{+\infty} n(x)\, dx=N$, where $N$ is the total number of
particles in the bound state. The density $n$ can be shown to be an even
function of $x$, and the matching condition at $x=0$ reads $n'(0^+)=2\,
\lambda\,n(0)$ ($\, '$ denotes $d/dx$). Using (\ref{bound1}) and these two
conditions we arrive at a set of two equations determining $n(0)$ and $\mu$ 
as functions of $\lambda$ and $N$,
\begin{equation}\label{bound2}
\frac{\lambda^2}{2}\,n(0)=\varepsilon(n(0))-\mu\,n(0)\; ,
\end{equation}
and
\begin{equation}\label{bound3}
\int_0^{n(0)} \frac{(n/2)^{1/2}\, dn}{\sqrt{\varepsilon(n)-\mu\,n}} = N \;
.\end{equation} For solving this system, one needs to know the explicit form
of the function $\varepsilon(n)$. We will see that for realistic values of
$\lambda$ corresponding to an attraction issued from a bend in the guide the
bound state can accommodate only a small number of particles. In the case of a
bend, it is thus sensible to concentrate on the low density limit (see the
estimate at the end of this section). Using (\ref{adia6a}), in the low density
regime we have $\varepsilon(n) = \epsilon_{0}\,n + a_{sc}\,n^2/a_\perp^2$ and
we obtain
\begin{equation}\label{bound4}
\mu=\epsilon_{0} - \frac{1}{2}\left( \int_{-\infty}^{+\infty} V_\parallel(x) \
 d x + \frac{a_{sc}}{a_\perp^2}\,N\right)^2 \; .
\end{equation} 
The first, negative, term inside the large parentheses is due to the attractive
potential produced by the bend (cf Eq.~(\ref{adia1a})), while the second one
comes from intra-atomic repulsive interactions in the condensate.
Eq.~(\ref{bound4}) clearly displays the existence of a threshold. When the
number of atoms $N$ occupying the bound state increases, the chemical
potential $\mu$ increases and eventually reaches the threshold $\epsilon_{0}$
at which the state disappears. This occurs for a number of atoms $N_{max}$
given by
\begin{equation}\label{bound5}
N_{max}=\frac{a^2_\perp}{a_{sc}}\, 
\left|\int_{-\infty}^{+\infty} V_\parallel(x) \ d x \right|
\; .\end{equation}

For an arbitrary potential, as $N\rightarrow N_{max}$ the spatial extension of
the bound-state diverges. Thus the approximation of $V_{\parallel}$ by a
$\delta$-function` is well justified in that limit, and we expect
Eq.~(\ref{bound5}) to be very accurate \cite{foot4}.

The order of magnitude of $N_{max}$ can be estimated by considering a bend of
constant radius of curvature $R_c$ and bending angle $\Theta$. From
(\ref{adia1a}) and (\ref{bound5}) we get $N_{max} =
a_\perp^2\Theta/(8a_{sc}R_c)$. For a guide with $R_c=5 a_\perp$,
$\Theta=\pi/2$ and $a_\perp$ ranging from 1 $\mu$m to 10 $\mu$m, $N_{max}$
ranges from 7 to 70 atoms for a condensate of $^{87}$Rb atoms ($a_{sc}=5.77$
nm). For $^{23}$Na, $N_{max}$ is doubled (since the s-wave scattering
length of $^{23}$Na is $a_{sc}=2.75$ nm).
	
If the attractive potential originates from a red detuned laser beam, using
the estimate of Sec.~\ref{experiments} (Eq.~(\ref{exp2})) one obtains $a_\perp
\int_{-\infty}^{+\infty} V_\parallel(x) \ d x \approx 10^{6}$ (to be compared
to the value $10^{-1}$ that applies to a bend). From this and from
Eq.~(\ref{bound5}) it follows that for rubidium $N_{max}$ can be as large as
$10^9$. In this case, however, one does not remain in the low density regime
where Eq.~(\ref{bound5}) holds (this regime is valid if $n(0)\,a_{sc}\ll 1$,
which from (\ref{bound2}) and (\ref{bound5}) gives $N_{max}\ll
a_\perp/a_{sc}\sim10^3$). Working in the high density regime instead, one
obtains
\begin{equation}\label{bound6}
N_{max}=\frac{3\,a_\perp^4}{16\,a_{sc}}\, \left|\int_{-\infty}^{+\infty}
V_\parallel(x) \ d x \right|^3 \; .
\end{equation}
One gets from this equation a value of $N_{max}$ of the order of $10^{20}$.
Note, however, that in this regime the high density approximation is valid at
$x=0$, but violated for large $x$ (when the density tends to zero). Hence,
without giving a precise order of magnitude, it is nevertheless clear from the
previous estimates that the maximum number of atoms the bound state can
accommodate is very large in this case.

\section{Transmission modes}\label{main}

From now on we concentrate on the stationary states of a beam of condensed
atoms connecting two reservoirs. For that purpose we look at the stationary
solutions (in the reference frame of the laboratory) of Eq.~(\ref{adia5}) with
$\psi$ having a finite value at $x\to\pm\infty$. We write
\begin{equation}\label{eq0}
\psi(x,t)=\exp\{-i\,\mu\,t\}\, A(x)\, \exp\{ i\varphi (x)\} \; ,
\end{equation}
with $A$ and $\varphi$ real functions. Since the wave function extends to
infinity, the chemical potential satisfies $\mu>\epsilon_{0}$. The density is
$n=A^2$, and the beam velocity is $v=\varphi'$. After factorizing out the
phases, Eq.~(\ref{adia5}) splits into two real equations corresponding to its
imaginary and real parts. The former imposes flux conservation, namely the
product $n(x)\,v(x)$ is a constant that we denote by $\Phi$,
$$
\Phi = n(x)\,v(x) \ .
$$
The real part gives a Schr\"odinger-like equation for $A(x)$
\begin{equation}\label{eq1}
- \frac{1}{2} A''+ \left[ \epsilon(n)+V_\parallel +\frac{\Phi^{2}}{2\, n^2}
 \right]\, A = \mu \, A \, . 
\end{equation}

In this section we consider the transmission modes of a condensate through a
straight wave guide with no obstacle. This corresponds to solving
Eq.~(\ref{eq1}) with $V_\parallel\equiv 0$. The modifications in the beam
density and phase produced by the presence of an obstacle are considered in
the following sections. When $V_\parallel\equiv 0$, Eq.~(\ref{eq1}) can be
integrated once, yielding
\begin{equation}\label{eq2}
\frac{1}{2} A^{\prime\,2} + W(n) = E_{cl} \quad\mbox{with}\qquad
W(n) = -\varepsilon(n) + \mu \, n +\frac{\Phi^2}{2\, n}\; .
\end{equation}
$\varepsilon(n)$ in (\ref{eq2}) is the same as in Eq.~(\ref{bound1}) and
$E_{cl}$ is an integration constant. This constant is denoted as a ``classical
energy'' because (\ref{eq2}) has an interpretation in terms of classical
dynamics. It expresses the energy conservation of a one-dimensional
Hamiltonian system for a fictitious classical particle with ``position'' $A$
and ``time'' $x$ moving in a potential $W(n=A^2)$, $E_{cl}$ being the total
energy of the particle. The solutions $A(x)$ therefore coincide with the
``classical'' solutions in the potential $W(n=A^2)$. The chemical potential
$\mu$ and the flux $\Phi$ fix the shape of the potential $W(n)$, while
$E_{cl}$ selects a ``trajectory'' $A(x)$ in this potential.

To clarify the physical meaning of $E_{cl}$ consider the linear (i.e.,
non-interacting) case $\varepsilon(n)=0$. Then the natural way to write the
solution (\ref{eq0}) is a superposition of plane waves,
\begin{equation}\label{eq0l}
\psi(x,t)=\exp\{-i\,\mu\,t\}\, \left[  \alpha \, \exp ( i k x ) + 
\beta \, \exp ( - i k x + i \theta ) \right] \; ,
\end{equation}
\noindent where $k$ is the wavevector and $\theta$ an arbitrary phase. In
terms of the two real parameters $\alpha$ and $\beta$, the flux and the
classical energy are written $\Phi = k \ (\alpha^2 - \beta^2)$ and $E_{cl} =
k^2 \ (\alpha^2 + \beta^2)$. $E_{cl}$ is therefore a measure of the total
intensity of the left and right incoming beams. $E_{cl}$ can be varied while
keeping $\Phi$ and $\mu$ constant (and therefore $W(n)$ constant) by changing
the amplitudes $\alpha$ and $\beta$ simultaneously while preserving the
difference $(\alpha^2 - \beta^2)$.

\begin{figure}[thb]
\begin{center}
\includegraphics*[width=7cm]{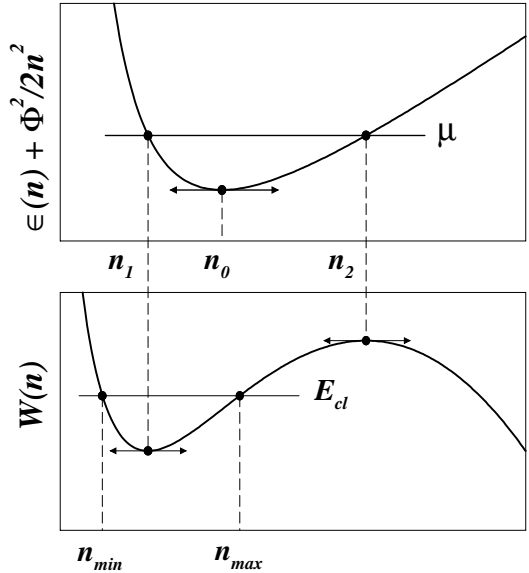}
\end{center}
\caption{\small Schematic behavior of the functions
$\epsilon(n)+\Phi^2/2\,n^2$ (top part) and $W(n)$ (bottom part) as a function
of $n$. $n_i$ ($i=1$ or 2) is defined by $\epsilon(n_i)+\Phi^2/2\,n_i^2=\mu$
and is a zero of $dW/dn$; $n_0$ is a zero of the second derivative
$d^2W/dn^2$. For $\mu$ and $\Phi$ given, a beam of uniform density has either
a density $n_1$ (and a velocity higher than the sound velocity $c(n_1)$) or a
density $n_2$ (and a velocity lower than $c(n_2)$). At a given $E_{cl}$,
$n_{min}$ and $n_{max}$ are the minimum and maximum values of the density
oscillations ($W(n_{min})=W(n_{max})=E_{cl}$).}
\label{schema}
\end{figure}

For studying the shape of $W(n)$ in Eq.~(\ref{eq2}) in the presence of
interactions it is customary to plot $\mu-dW/dn=\epsilon(n)+\Phi^2/(2 n^2)$ as
a function of $n$, as represented in the top part of Fig.~\ref{schema}. At low
densities ($n \rightarrow 0$) the term $\Phi^2/(2n^2)$ dominates. At high
densities the intra-atomic interaction contained in $\epsilon (n)$ takes over,
and leads to a monotonic growth for large values of $n$ (due to the repulsive
interactions, $\epsilon(n)$ is an increasing function of $n$). At intermediate
values there is a minimum at a density denoted $n_0$. The relevant case
(leading to finite densities at infinity) corresponds to $\mu \ge
\epsilon(n_0)+\Phi^2/(2n_0^2)$, and is shown in Fig.~\ref{schema}. Then the
derivative of $W(n)$ is zero for two densities $n_1$ and $n_2$, with $n_1\le
n_0\le n_2$. The corresponding plot of the potential $W(n)$ is shown in the
lower part of the figure. In order to have a finite density at infinity one
should also impose a bounded motion of the fictitious classical particle, and
this results in the two additional conditions (i) $W(n_1)\le E_{cl} \le
W(n_2)$, and (ii) $n(x) \leq n_2$ for any value of $x$.

The different types of solution $A(x)$ are therefore described by the
different motions a classical particle undergoes in a potential $W(n)$ at the
allowed energies $E_{cl}$. The two simplest solutions corres\-pond to the
fixed points of the potential, $n(x)=n_1$ or $n_2$, where the ``classical''
particle remains at rest. They correspond to constant density solutions. Since
the densities are different ($n_2 > n_1$) and the flux has the same value
$\Phi$, the velocities $v_j$ ($j=1$ or 2) of the condensed beam are also
different, with $v_2 < v_1$. $v_1$ (resp. $v_2$) corresponds to a beam
velocity above (resp. below) the speed of sound. To see this, we first note
(see below) that for a condensate {\it at rest} (i.e., $v\equiv 0$) with
uniform density $n(x)=n$, the sound velocity $c$ is defined by
\begin{equation}\label{sound}
c^2(n)=n\,\frac{d\epsilon}{dn} \; .
\end{equation}
In the case of a moving condensate with uniform density $n$, one has a well
defined velocity $v=\Phi/n$ given by $\mu=\epsilon(n)+\frac{1}{2}v^2$ (see
Eq.~(\ref{eq1})). From Eqs. (\ref{eq2}) and (\ref{sound}), we have moreover
$d^2W/dn^2=(v^2-c^2(n))/n$. Since $d^2W/dn^2$ at $n_1$ (resp. $n_2$) is
positive (resp. negative), it follows that $v_1 > c(n_1)$ (resp. $v_2 <
c(n_2)$).

We now consider the density profile of the transmission modes in the vicinity
of the constant solution $n(x)=n_1$. For energies $E_{cl}$ slightly higher
than $W(n_1)$ the stationary solutions are sinusoidal waves of the form
\begin{equation}\label{sinus}
A(x) = A_1 + \frac{\sqrt{2}}{k} [E_{cl}-W(n_1)]^{1/2}  
\cos \left( k \ x + \theta \right) \; ,
\end{equation}
\noindent where $k^2=\left.d^2W/dA^2\right|_{A_1} = 4(\Phi^2/n_1^2 - c^2
(n_1))$ and $n_1 = A_1^2$ is the constant density. This sinusoidal wave is
reminiscent of a sound wave, but sound waves are progressive whereas
Eq.~(\ref{sinus}) describes a standing wave. The structure of the sound wave
(and the whole spectrum of elementary excitations) is better displayed by a
slight modification of the procedure used so far. Instead of the stationary
ansatz (\ref{eq0}) one looks for solutions of the form $\psi(x,t)=\exp\{-i\mu
t\} A(x-u\,t) \exp\{i\phi(x-u\,t)\}$, where $u$ is a constant parameter,
physically interpreted as the velocity of an arbitrary moving frame. Mass
conservation now reads $\Phi=n(v-u)$ ($\Phi$ is the flux in the moving frame)
and Eqs.~(\ref{eq1},\ref{eq2}) keep the same form, with $\mu$ replaced by
$\mu+u^2/2$, all functions now depending on $X=x-u\,t$ and not merely on $x$.
The constant solution $n(x)=n_1$ can now be given zero velocity ($v_1=0$) if
one chooses $u$ such that $u+\Phi/n_1=0$ (this is of importance because one
wishes to study elementary excitations in a system at rest). In this case, a
perturbative treatment of (\ref{eq2}) for $E_{cl}$ near $W(n_1)$ again gives a
solution of the form (\ref{sinus}) with $x$ replaced by $X$. This is a
progressive wave depending on $k\,X=k\,x-\omega_k\,t$ with $\omega_k = k\,u$,
and thus satisfying the Bogoliubov dispersion relation
\begin{equation}\label{eq3}
\omega_k^2 = k^2 \left( n_1\left.\frac{d\epsilon}{dn}\right|_{n_1} +
\frac{k^2}{4} \right) \; .
\end{equation}
The long wave-length limit of (\ref{eq3}) corresponds indeed to sound waves
with a sound velocity $c(n_1)$ as given by Eq.~(\ref{sound}). It is also
possible to obtain in this way the dispersion relation of the elementary
excitations of a beam moving at constant velocity $v_1$ (which is simply
Eq.~(\ref{eq3}) Doppler shifted).

Note that our approach is unable to reproduce the decrease of slope of the
spectrum of elementary excitations that occurs in the high density regime, for
wave vectors $k$ of the order of the transverse extension $R_\perp$ of the
condensate. This effect, predicted in Refs.~\cite{Zar98,Str98} and observed
numerically in \cite{Fed00} goes beyond the quasi-1D approach: it occurs when
the excitation has a wavelength allowing exploration of side regions of the
condensate that have lower local sound velocity. Hence, it cannot be
reproduced by using the adiabatic ansatz (\ref{adia1}).

\begin{figure}[thb]
\begin{center}
\includegraphics*[width=7cm]{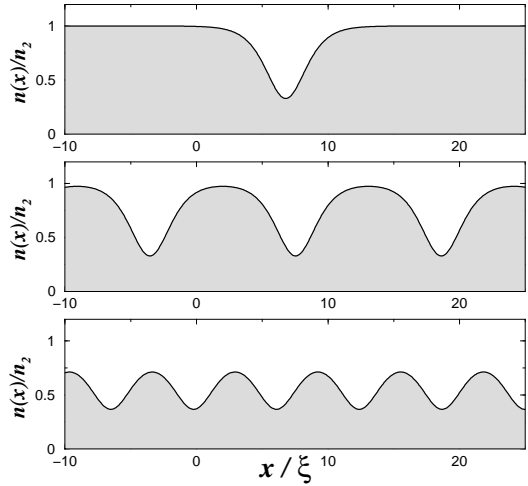}
\end{center}
\caption{\small Beam density along a straight guide (high density regime) with
$v_2/c_2=0.7$ (lengths are given in units of $\xi (n_2)$, see
Eq.~(\ref{adia7})). Top: solitary wave ($E_{cl} = W(n_2)$). Middle: cnoidal
waves existing for $W(n_1) < E_{cl} < W(n_2)$. Bottom: for $E_{cl} \gsim
W(n_1)$ the cnoidal wave deforms continuously to a sinusoidal small-amplitude
wave.}
\label{cnoidales}
\end{figure}

The structure of the stationary solutions for energies $E_{cl}$ close to (but
lower than) $W(n_2)$ is totally different from the sinusoidal waves we just
discussed (which exist for $E_{cl} \gsim W(n_1)$). The uniform solution $n(x)
= n_2$ coexists with a solitary wave corresponding in the classical analogy to
a motion along the separatrix located at $E_{cl}=W(n_2)$. This solitary wave
has constant density $n_2$ at $x\to\pm\infty$, and a trough whose minimum
density satisfies the condition $W(n_{min}) = W(n_2)$ (cf.
Fig.~\ref{cnoidales}, top part). As the energy is lowered from
$E_{cl}=W(n_2)$, density oscillations appear whose amplitude decreases as
$E_{cl}$ diminishes. These solutions are cnoidal waves (see, e.g.,
\cite{Whi74}) with periodic oscillation between two values $n_{min}$ and
$n_{max}$, as defined in Fig.~\ref{schema}. As the energy $E_{cl}$ is further
reduced the density profile tends continuously to the sinusoidal waves
discussed above. This transition is illustrated in Fig.~\ref{cnoidales}, which
is drawn in the high density regime for a transverse harmonic confining
potential (i.e., with $\epsilon(n)$ given by Eq.~(\ref{adia6b})). The same
qualitative behavior is valid for any density regime.

Figure~\ref{cnoidales} summarizes the possible density profiles of the
transmission modes of the condensate along a straight guide. In the remaining
sections we consider the modifications induced by the presence of an obstacle
in the flow of the condensate. Finding the transmission modes now reduces to a
scattering problem in which two of the ``free'' modes discussed in this section
are matched by the potential representing the obstacle. The correct boundary
conditions to be imposed are determined by the relative value of the phase
velocity
\begin{equation}\label{pv}
v_p(k) = \frac{\omega_k}{k}
\end{equation}
with respect to the group velocity
\begin{equation}\label{gv}
v_g(k) = \frac{\partial \omega_k}{\partial k} 
= v_p(k) \left( 1 + \frac{k^2}{k^2+4\,c^2(n)} \right)  \ .
\end{equation}
Both functions start from the value $c (n)$ at $k = 0$ and then increase
monotonically, with $v_g (k) > v_p (k)$ for any $k>0$. For the stationary
motion of an obstacle in a condensate at rest, $v_p$ coincides with the
velocity of the obstacle with respect to the beam, this is the condition of
stationarity. The energy transfered to the fluid propagates with a velocity
$v_g$ greater than the velocity of the obstacle with respect to the beam
(since $v_g>v_p$). As a consequence, radiation conditions require that the wake
is always located {\it ahead} of the obstacle (i.e., up-stream in a frame
where the disturbance is at rest), with no long-range perturbation of the
fluid on the down-stream side \cite{Lamb}.

\section{The attractive square well}\label{asw}

Having recalled the different solutions existing in a straight wave guide, we
now discuss the influence on the transmission modes of a longitudinal
potential representing a motionless obstacle placed in the trajectory of the
beam. Specifically, we consider a potential $V_\parallel(x)$ that vanishes
everywhere except in a finite region $0\le x\le\sigma$ where it takes the
constant value $-V_0$ ($V_0>0$). If its origin is the presence of a bend, the
square well potential corresponds to a wave guide with a constant curvature
over a finite length $0 < x < \sigma$ and straight elsewhere \cite{refadia}.
Apart from considerations related to its physical origin, this model potential
is of interest because it allows one to understand in a simple case the
different stationary regimes occurring also in more complex potentials.

An important point for the determination of the transmission modes of the
condensate along the guide is the boundary conditions. As discussed at the end
of the previous section, amongst all the possible stationary solutions that
exist in the presence of a scattering potential, the only physical ones are
those that tend to a flat density down-stream. Hence we consider density
profiles tending to a flat density at $x \to -\infty$, with $n(x \to -\infty)
\to n_\infty$, and with a {\it negative} velocity $v_\infty=\Phi/n_\infty$.
The sound velocity at infinity $c_\infty=c(n_\infty)$ will also be chosen
negative in all the following. This corresponds to a beam incoming from the
right, unperturbed far down-stream by the presence of $V_\parallel$, and
characterized by the two parameters $v_\infty$ and $n_\infty$ (or equivalently
by $\mu$ and $\Phi$, since $\mu=\epsilon(n_\infty)+v_\infty^2/2$ and $\Phi =
n_\infty\,v_\infty$). Moreover, we will systematically express lengths in
units of $\xi = [2 (\epsilon (n_\infty) - \epsilon_0)]^{-1/2}$ (cf
Eq.~(\ref{adia7})).

When $V_\parallel$ is a square well, Eq.~(\ref{eq1}) takes a particularly
simple form. Everything happens as for a straight wave guide, except that $\mu$
in (\ref{eq1}) is shifted to $\mu+V_0$ in the region $0\le x\le\sigma$. Hence,
as in the case of a straight guide, one has an integral of motion, but it
takes a different value in each portion of space,
\begin{equation}\label{asw1}
\left\{
\begin{array}{rcc}
\frac{1}{2}\,A^{\prime\,2} + W(n) = E_{cl}^- & 
\;\;\;\; & x\le 0 \; , \\
 & & \\
\frac{1}{2}\,A^{\prime\,2} +W(n)+V_0\,n=E_{cl}^0 &
 & 0\le x\le\sigma\; , \\
 & & \\
\frac{1}{2}\,A^{\prime\,2} + W(n) = E_{cl}^+ &
 & \sigma\le x\; . \\
\end{array}
\right.
\end{equation}
$W(n)$ in Eq.~(\ref{asw1}) is defined as in Eq.~(\ref{eq2}). $E_{cl}^-$,
$E_{cl}^0$ and $E_{cl}^+$ are the values of the integration constant in each
region. Since the solution is flat far down-stream (when $x\to-\infty$) one has
$E_{cl}^- = W(n_\infty)$. The matching at $x=0$ and $x=\sigma$ imposes
continuity of the density and of its derivative. This leads to
\begin{equation}\label{asw2}
E_{cl}^- + V_0\,n(0) = E_{cl}^0 = E_{cl}^+ + V_0\,n(\sigma) \; .
\end{equation}
Different types of solution satisfying Eqs.~(\ref{asw1}) exist, depending
whether the far down-stream beam velocity $v_\infty$ is greater or smaller
than the speed of sound $c(n_\infty)=c_\infty$. We consider these two
different regimes separately.

\subsection{low beam velocity: $v_\infty / c_\infty < 1$} 

The first type of solution we consider is rather intuitive if employing a
perturbative treatment (see Sec. \ref{simple}). It corresponds to solutions
with an increased density in the region of the potential. In the following, we
refer to these solutions as ``B-solutions'' (where B stands for bump).

\begin{figure}[thb]
\begin{center}
\includegraphics*[width=10cm]{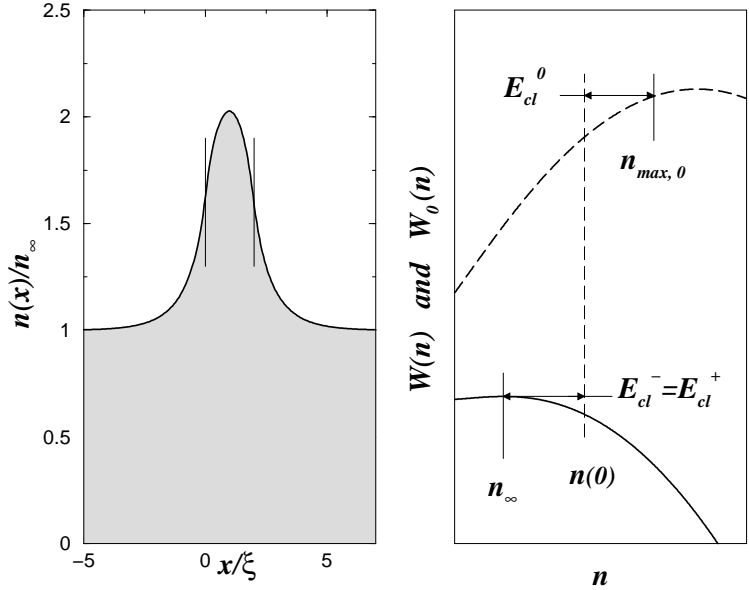}
\end{center}
\caption{\small Type-B stationary solution in an attractive square well. The
left plot displays the density profile. The two vertical solid lines indicate
the location of the square well. This plot corresponds to $\sigma=2\,\xi$ and
$V_0\,\xi^2=0.5$ in a low-density beam at velocity $v_\infty/c_\infty=0.7$
(the healing length $\xi$ is defined as in (\ref{adia7}) and is computed at
density $n_\infty$). The right plot illustrates the behavior of the solution
in the diagram $(n,W(n))$. The solid curve represents $W(n)$ and the dashed
one $W_0(n)$. In the rear of the obstacle, the density evolves from $n_\infty$
to $n(0)$ in the potential $W(n)$ with a classical energy $E_{cl}^- =
W(n_\infty)$, then from $n(0)$ to $n_{max,0}$ and back to $n(0)$ in the
potential $W_0(n)$ (with energy $E_{cl}^0$), and finally the up-stream density
goes from $n(0)$ to $n_\infty$ in $W(n)$.}
\label{bump1}
\end{figure}

The B-solutions are found by looking for solutions with a density increasing
when $x$ moves from $x=-\infty$ towards the origin. Since at $x\rightarrow
-\infty$ the density has the constant value $n_\infty$, and since we are
imposing $v_\infty / c_\infty < 1$, $n_\infty$ coincides with the uniform
density denoted $n_2$ in Sec.~\ref{main}, and $E_{cl}^-=W(n_\infty)$. But
contrary to the solitary wave of Sec.~\ref{main} in which the density
decreases, in a type-B solution we move to the right of $n_\infty = n_2$ along
the separatrix (see Fig.~\ref{bump1}) and the B-solution has a density peak
instead of a trough.

	Since the boundary conditions are fixed down-stream, here and in all the
following, we find it more convenient to integrate Eqs.~(\ref{asw1}) starting
far in the rear of the obstacle, i.e. from left infinity (remember that the
beam is incident from the right). Starting from a value $n_\infty$ at
$x=-\infty$, the density has reached a value $n(0)$ at $x=0$. From this point
on the equivalent ``classical particle'' moves in the potential
\begin{equation}\label{w0at}
W_0(n)=W(n)+V_0\,n
\end{equation} 
at an energy $E_{cl}^0 = E_{cl}^{-} + V_0 n(0)$. At $x=\sigma$ it evolves
again in $W(n)$. Since to reach a finite density when $x\to+\infty$ one
needs to have $E_{cl}^+\le W(n_\infty)=E_{cl}^-$, from (\ref{asw2}) this
imposes $n(\sigma)\ge n(0)$. But, if the inequality is strict, one has
$E_{cl}^+<W(n_\infty)$ and $n(\sigma)>n(0)\ge n_\infty$: this case should be
excluded since from Fig.~\ref{schema} one sees that this leads to a diverging
density at right infinity (the classical particle escapes to infinity). Hence
one should have $n(0)=n(\sigma)$, $E_{cl}^+=E_{cl}^-$ and the B-solutions are
even.

These solutions exist for any type of attractive square well. For a given beam
(characterized by $n_\infty$ and $v_\infty$) and given values of $\sigma$ and
$V_0$, the value of $A(0)$ is determined by demanding that the amplitude
varies from $A(0)$ to its maximum value $A_{max,0}$ and back over a distance
$\sigma$. $A_{max,0}$ is determined as a function of $A(0)$ from the equation
$W_0(A)=E_{cl}^0$ whose two smallest positive solutions are denoted
$A_{min,0}$ and $A_{max,0}$ (in all the following we denote with an index
``$0$'' the quantities concerning $W_0$ and defined as for $W$ in
Fig.~\ref{schema}). We have
\begin{equation}\label{asw3}
\sigma=2\, \int_{A(0)}^{A_{max,0}} \frac{dA}{A'}
=\sqrt{2}\, \int_{A(0)}^{A_{max,0}} 
\frac{dA}{\sqrt{E_{cl}^0-W_0(A)}} \; .
\end{equation}

\begin{figure}[thb]
\begin{center}
\includegraphics*[width=10cm]{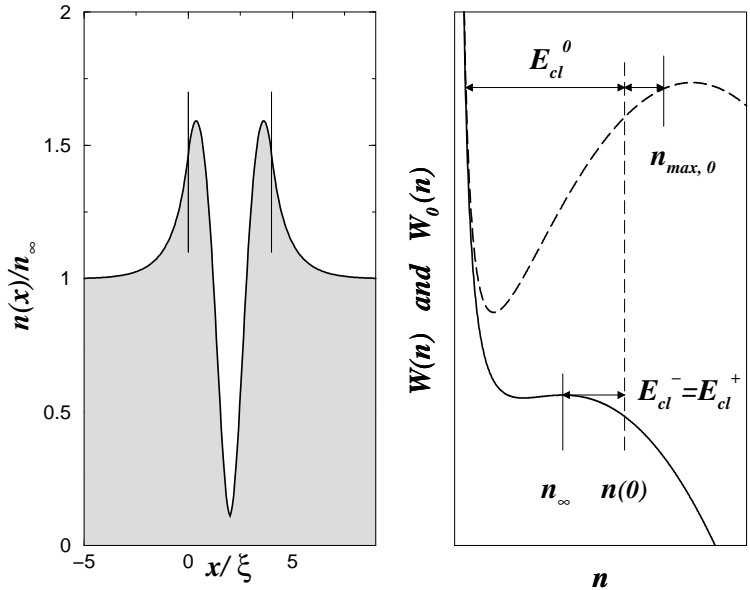}
\end{center}
\caption{\small A B$_2$ solution in an attractive square well. Left and right
parts as in Fig.~\ref{bump1}. The plot corresponds to $\sigma=4\,\xi$ and
$V_0\,\xi^2=0.5$ in a low density beam at velocity $v_\infty/c_\infty=0.7$.
The down-stream density evolves from $n_\infty$ to $n(0)$ in the potential
$W(n)$ with a classical energy $E_{cl}^- = W(n_\infty)$, then it makes a
complete oscillation in $W_0(n)$ starting from $n(0)$ towards $n_{max,0}$ and
ending again at $n(\sigma) = n(0)$. Finally the up-stream density goes from
$n(\sigma)$ to $n_\infty$ in the potential $W(n)$.}
\label{bump2}
\end{figure}

For sufficiently small $\sigma$ the only existing B-solution is the one
described above. However, new B-solutions appear as the width $\sigma$
increases, because the ``classical particle'' before evolving back in $W(n)$
has enough ``time'' to make one (or several) oscillations in $W_0 (n)$. The
general density profile of the B-type increases from $n_\infty$ at $x=
-\infty$ up to $x=0$, and has $N$ maxima and $N-1$ minima between $x=0$ and
$x=\sigma$, with $N=1,2,3..$. We denote this a B$_{\sss N}$-solution. Figure
\ref{bump1} corresponds to a B$_1$ solution. The behavior of a B$_2$ solution
is illustrated in Fig.~\ref{bump2}. For an arbitrary $N$, Eq.~(\ref{asw3})
takes the form
\begin{equation}\label{asw4}
\sigma=2\, \int_{A(0)}^{A_{max,0}} \frac{dA}{A'} +
2\,(N-1)\, \int_{A_{min,0}}^{A_{max,0}} \frac{dA}{A'} \; ,
\end{equation}
\noindent where $A_{min,0}$ and $A_{max,0}$ are, as indicated before, the two
smallest positive solutions of $W_0(A)=E_{cl}^0$. The width of the potential
below which a given solution disappears occurs when
$A(0)=\sqrt{n_\infty}=A_{max,0}$. In this case the solution is perfectly flat
for $x\le 0$ and $x\ge\sigma$ and is a portion of cnoidal wave (with $N$
oscillations) in the region $0\le x\le\sigma$. For a given $V_0$, this forces
$\sigma$ to be larger than the value $\sigma_{\sss N}(V_0)$,
\begin{equation}\label{asw5}
\sigma\ge \sigma_{\sss N}(V_0)=\sqrt{2}\, (N-1)\,
\int_{\sqrt{n_\infty}}^{A_{max,0}} 
\frac{dA}{\sqrt{E_{cl}^0-W_0(A)}}
\; .\end{equation}

Other types of solution, different from the B-family, exist for a beam
velocity lower than the speed of sound. They correspond to density profiles
that {\it decrease} from the down-stream asymptotic value $n_\infty$ as $x$
moves from $-\infty$ towards the origin, with $E_{cl}^-=W(n_\infty)$. Hence
this type of solution is a portion of a solitary wave in the rear of the
obstacle. We refer to these solutions as the D-solutions (where D stands for
depressed).

\begin{figure}[thb]
\begin{center}
\includegraphics*[width=10cm]{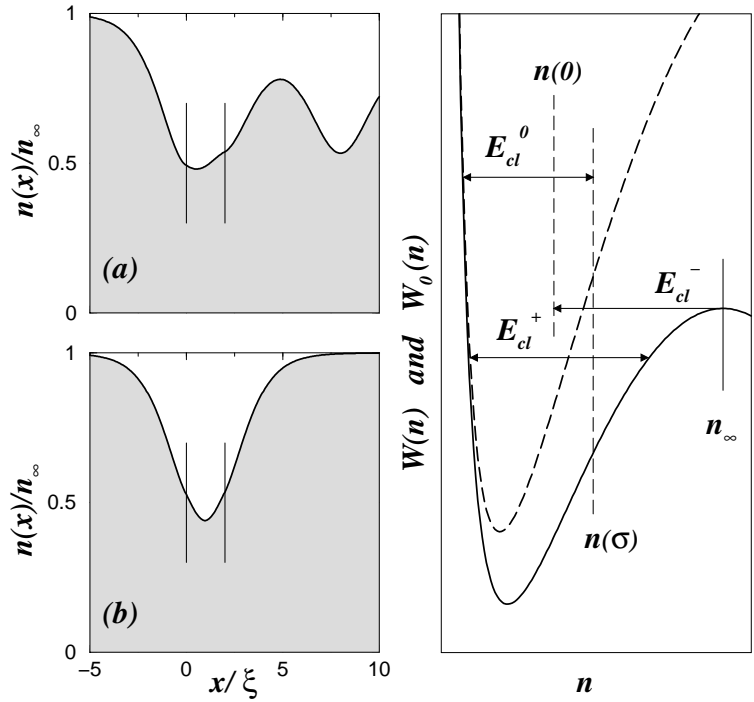}
\end{center}
\caption{\small A D$_1$ solution in an attractive square well. Left and right
parts as in Fig.~\ref{bump1}. Plot (a) is the generic case (with
$n(\sigma)>n(0)$) and (b) is the symmetric case where $n(\sigma)=n(0)$. These
plots correspond to $\sigma=2\,\xi$ and $V_0\,\xi^2=0.1$ in a low density beam
at velocity $v_\infty/c_\infty=0.7$. The right plot illustrates the behavior
of the typical solution (such as displayed in part (a)) in the diagram
$(n,W(n))$. The down-stream density evolves from $n_\infty$ to $n(0)$ in the
potential $W(n)$ with a classical energy $E_{cl}^- = W(n_\infty)$. Then in
$W_0$ (with an energy $E_{cl}^0$) from $n(0)$ to the minimum density
$n_{min,0}$ and back to a value $n(\sigma)$ (larger than $n(0)$). Finally the
up-stream density oscillates in the classical potential $W(n)$ as a cnoidal
wave.}
\label{hole1}
\end{figure}

For a D-solution, the density has a value $n(0)<n_\infty$ at $x=0$ and from
there on, the equivalent classical particle evolves in the potential $W_0$. In
the simplest case the particle bounces once on the repulsive core at the
origin, namely the density further decreases until it reaches a value
$n_{min,0}$ (satisfying $E_{cl}^0=W_0(n_{min,0})$) and then increases until
$x=\sigma$. Then the classical particle evolves in $W(n)$ again, with and
energy $E_{cl}^+$ that has to be lower than or equal to $W(n_\infty)$, and
this imposes $n(\sigma)\ge n(0)$ (cf Eq.~(\ref{asw2})). Note that here,
contrary to the case of the B-solution, the strict inequality is possible; it
corresponds to $E_{cl}^+<W(n_\infty)$ and $n(\sigma)<n_\infty$, i.e., the
up-stream solution is a cnoidal wave. In the particular case that
$n(\sigma)=n(0)$ the solution is even. A generic density profile with a
cnoidal wave is represented in Fig.~\ref{hole1}(a), and the even solution is
represented in Fig.~\ref{hole1}(b).

For a given well depth $V_0$ the simple D-solutions of Fig.~\ref{hole1} (which
we denote as the D$_1$ solution, with one minimum in the region of the
potential) do not exist for all values of $\sigma$. When the well becomes very
large, $\sigma$ may exceed the period of the oscillation of
density in the well (i.e., the ``time'' period of the ``classical particle''
evolving in $W_0$). In that case the D$_1$-solution disappears. The limiting
case corresponds to $n(0)=n_\infty=n(\sigma)$ i.e., to a flat density outside
the region of the well having one oscillation in the region of the well. This
upper limit is exactly the lower limit $\sigma_2(V_0)$ below which the B$_2$
solution does not exist. As a consequence, when $\sigma$ increases from a
small value the D$_1$-solution disappears at $\sigma=\sigma_2(V_0)$ and
becomes the B$_2$-solution (of the type illustrated in Fig.~\ref{bump2}),
which is then allowed for any larger value of $\sigma$. The point is that for
a B-solution, when changing from potential $W$ to $W_0$ in the $(n,W)$
diagram, one can jump arbitrarily close to the separatrix of $W_0$, thus
making the period in the region of the potential as large as desired. Hence,
once a B-solution exists for a given $\sigma$, it exists also for any larger
value. For a D-solution however the period in $W_0$ is limited: it takes its
largest value if one enters and leaves the region of the potential $W_0$ with
a flat density (i.e., $n(0)=n_\infty=n(\sigma)$).

\begin{figure}[thb]
\begin{center}
\includegraphics*[width=10cm]{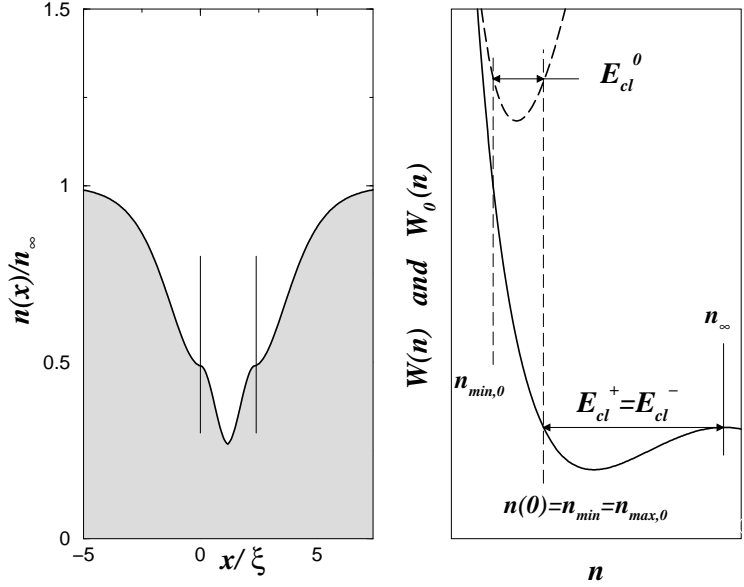}
\end{center}
\caption{\small The D$_1$ solution in a strongly attractive square well of
width just above $\sigma_1^*(V_0)$. Left and right parts as in
Fig.~\ref{bump1}. The density outside the well is composed of two half
solitons. This plot corresponds to $V_0\,\xi^2=0.5$ and $\sigma\simeq
2.4\,\xi$ in a low density beam at velocity $v_\infty/c_\infty=0.7$. The
density evolves from $n_\infty$ (far down-stream) to $n(0) = n_{min}$ in the
potential $W(n)$ with classical energy $E_{cl}^- = W(n_\infty)$. It then
evolves in $W_0$ from $n(0)$ to $n_{min,0}$ and back. Finally the up-stream
density oscillates in the classical potential $W(n)$ from $n(\sigma)=n_{min}$
back to $n_\infty$.}
\label{hole1bis}
\end{figure}

The existence of the D$_1$ solution for $\sigma<\sigma_2(V_0)$ depends on the
value of $V_0$ and of the relative positions of the curves $W(n)$ and $W_0(n)$
in the $(n,W)$ diagram. One regime is set by small values of $V_0$ such that
the condition $n_{min}<n_{1,0}$ is satisfied, where $n_{min}$ is the smallest
positive solution of $W(n)=W(n_\infty)$ (see Fig.~\ref{schema} in the case
$E_{cl}=W(n_2=n_\infty)$) and $n_{1,0}$ is the first zero of $dW_0/dn$
(remember that we denote with an index ``$0$'' the quantities concerning $W_0$
and defined as in Fig.~\ref{schema} for $W$). In this case one can easily
check that any value of $\sigma$ smaller than $\sigma<\sigma_2(V_0)$
corresponds to an acceptable B$_1$ solution. On the contrary, for larger
values of $V_0$ for which $n_{min}>n_{1,0}$ is satisfied there is a minimum
width $\sigma_1^*(V_0)$ below which D$_1$ solutions do not exist:
\begin{equation}\label{asw6}
\sigma_1^*(V_0)=\sqrt{2}\,\int_{A_{min,0}}^{A_{max,0}}
\frac{dx}{\sqrt{E_{cl}^0-W_0(n)}} \; ,
\end{equation}
where $E_{cl}^0=W(n_{min})+V_0\,n_{min}$, $A_{min,0}$ and $A_{max,0}$
being solutions of $W_0(A)=E_{cl}^0$ (one has
$A_{max,0}=A_{min}$). The limiting
case $\sigma=\sigma_1^*(V_0)$ is illustrated in Fig.~\ref{hole1bis}.

\begin{figure}[thb]
\begin{center}
\includegraphics*[width=10cm]{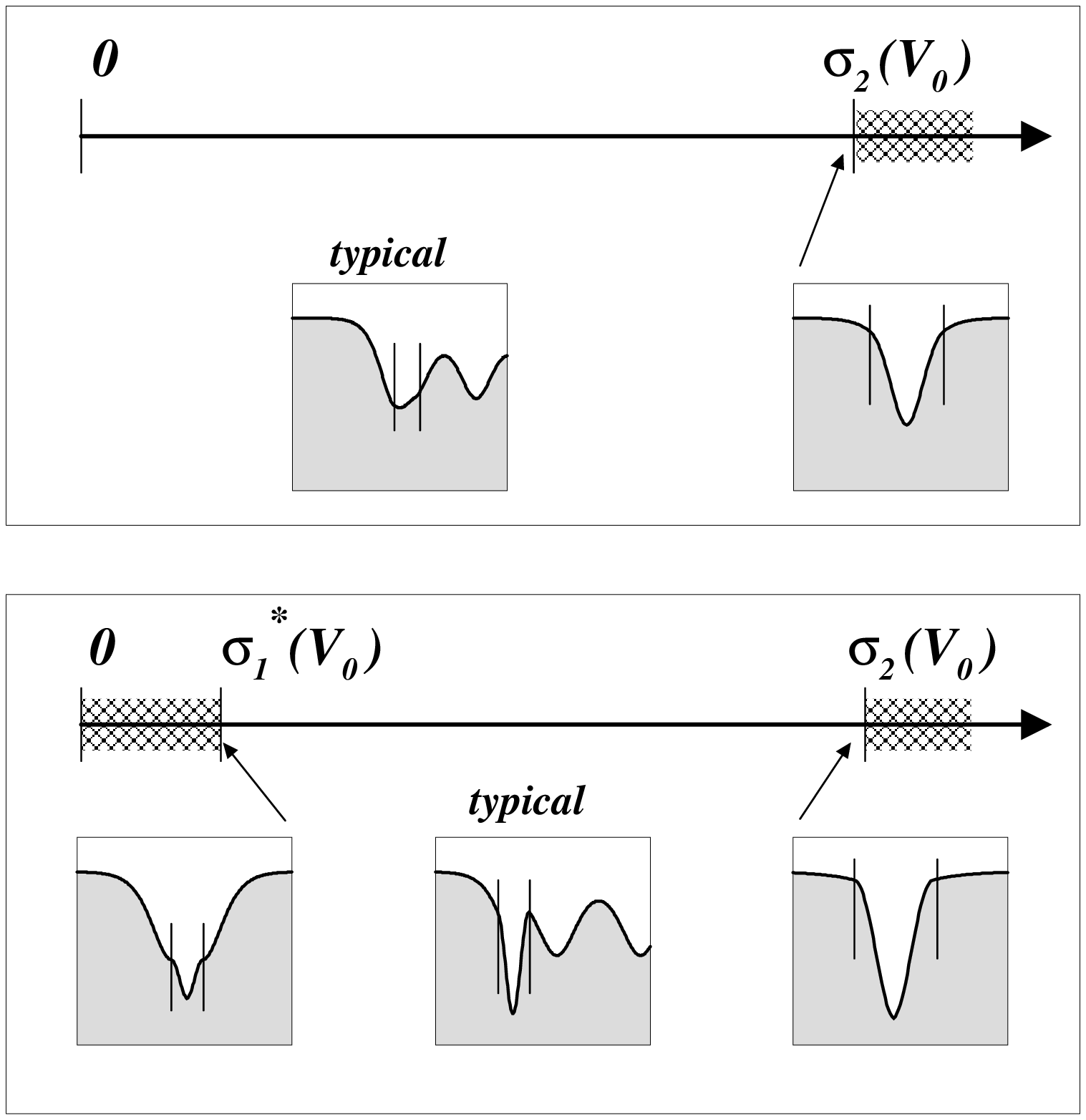}
\end{center}
\caption{\small Synoptic diagram of the evolution of the morphology of the
D$_1$ solution for a given $V_0$, as $\sigma$ increases. The insets display
the density profiles $n(x)/n_\infty$. Hatched regions indicate values of
$\sigma$ for which the solution does not exist. For shallow potentials (for
which $n_{1,0}>n_{min}$, upper part of the plot) the D$_1$ solution exists for
any width $\sigma$ between 0 and $\sigma_2(V_0)$. For deep potentials
($n_{1,0} < n_{min}$, lower part) it exists only if
$\sigma\in[\sigma_1^*(V_0),\sigma_2(V_0)]$. In both cases (shallow or deep
potential) at $\sigma=\sigma_2(V_0)$ the D$_1$ solution disappears and
becomes a B$_2$ solution.}
\label{table1}
\end{figure}

The situation for the solutions of type D$_1$ is summarized in Figure
\ref{table1}.

\begin{figure}[thb]
\begin{center}
\includegraphics*[width=8cm]{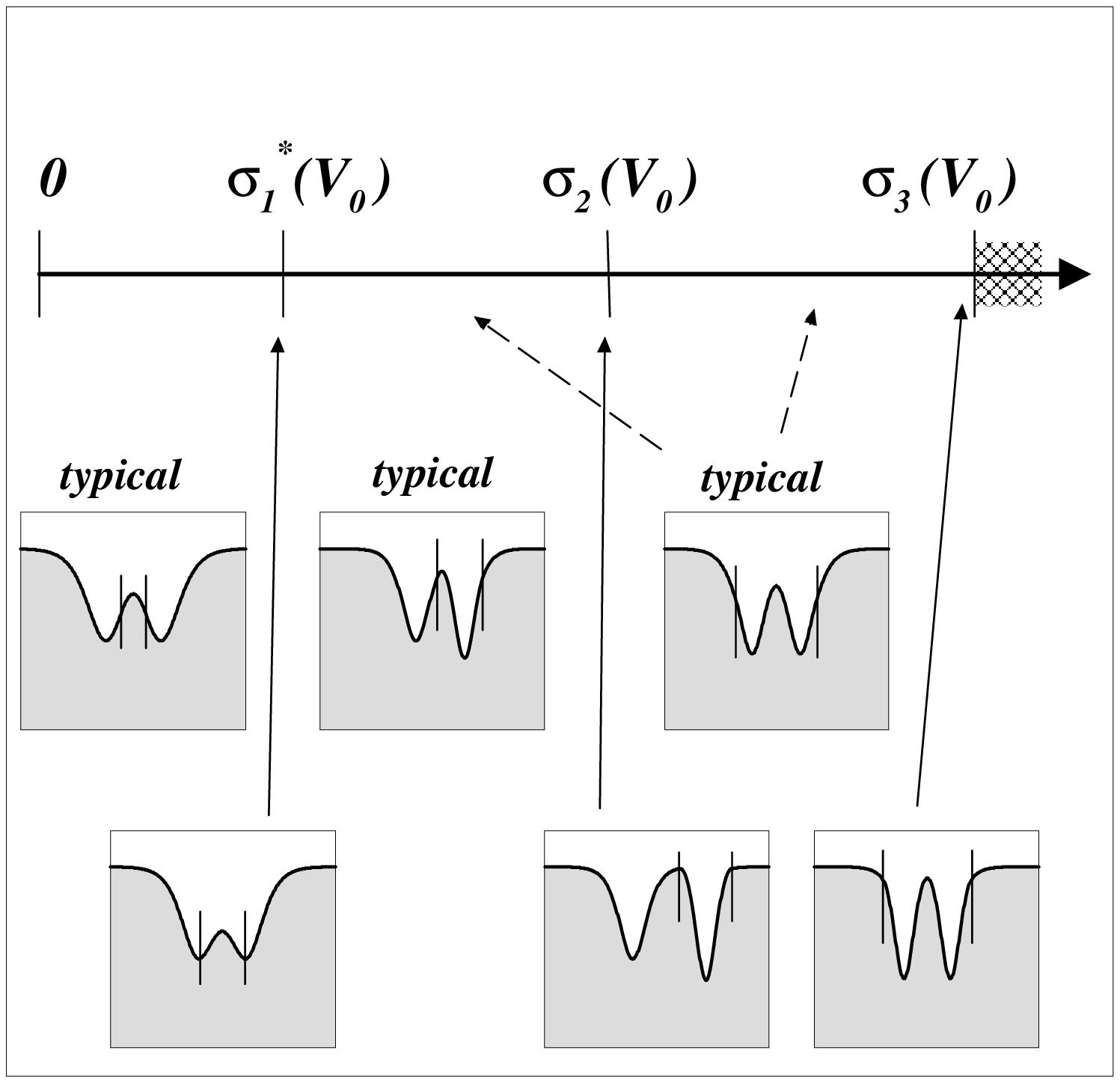}
\includegraphics*[width=8cm]{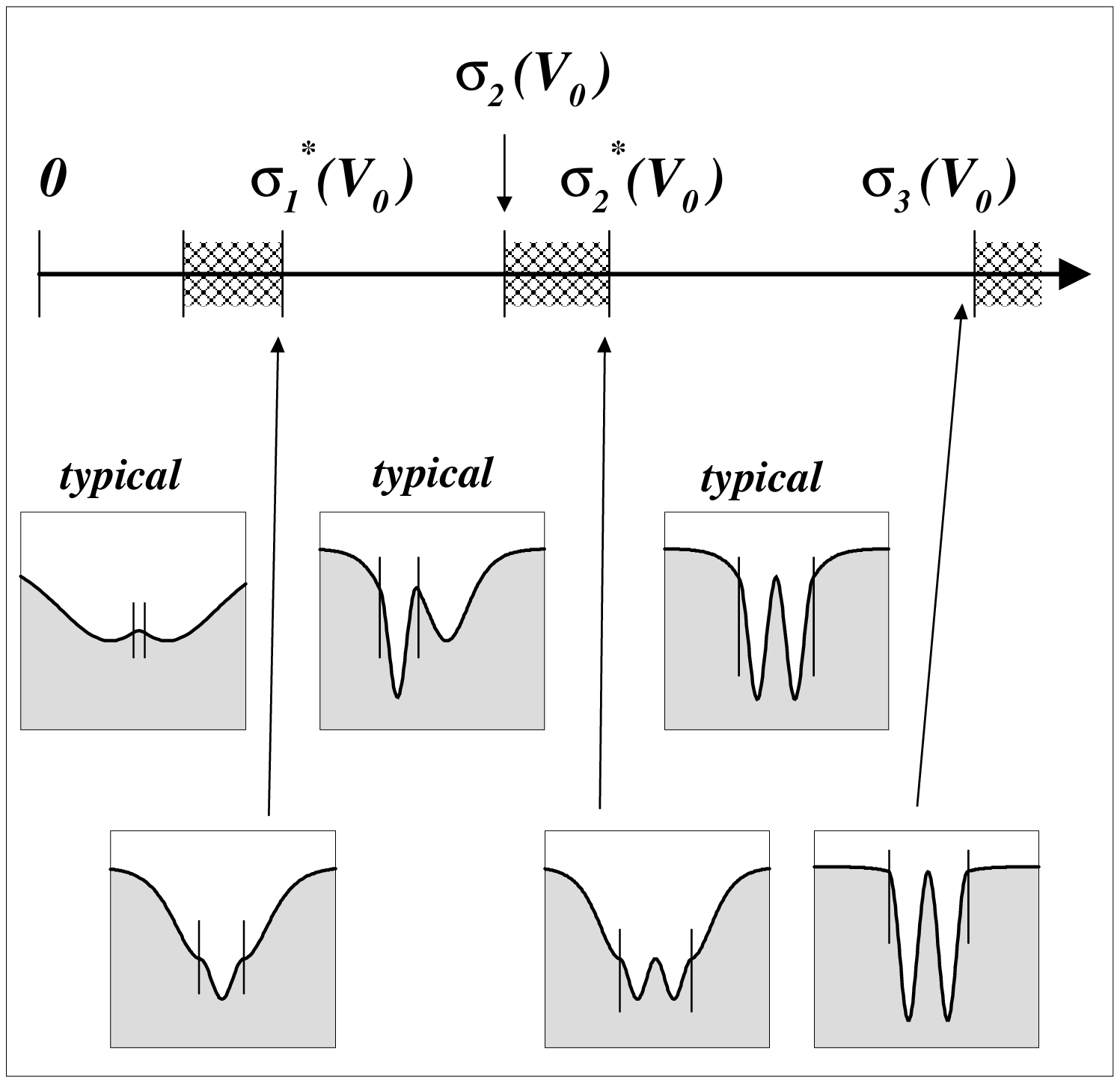}
\end{center}
\caption{\small Synoptic diagram of the evolution of the morphology of the
D$_2$ solution for given $V_0$, as $\sigma$ increases (see caption of
Fig.~\ref{table1}). For $\sigma$ larger than $\sigma_3(V_0)$ the D$_2$
solution disappears (it becomes a B$_3$ solution). The left part corresponds
to a shallow well and the right part to a deep well. For legibility, the cases
denoted as ``typical'' have been taken to consist ahead of the obstacle (for
$x\ge\sigma$) of a portion of a soliton (whereas the most generic solutions
are cnoidal waves).}
\label{table2}
\end{figure}

D-solutions can oscillate in the region of the well, as B-solutions do. We
denote by D$_{\sss N}$-solution a type-D solution with $N$ minima. Contrary to
the case of B$_{\sss N}$-solutions, there exists a maximum width for D$_{\sss
N}$-solutions to occur. When the D$_{\sss N}$-solution disappears, it becomes
a B$_{\sss N+1}$-solution (exactly as discussed above for $N=1$). D$_{\sss N}$
solutions are not necessarily even as B$_{\sss N}$ solutions are; ahead of the
obstacle (for $x \geq \sigma$) they typically consist of a cnoidal wave (see
the discussion for $N=1$ and Figure \ref{hole1}). In this case we do not
consider that the minima of the cnoidal wave outside the well increase the
index ``$N$'' in the name ``D$_{\sss N}$-solution'' (for instance the profile
of Fig.~\ref{hole1}(a) corresponds to a D$_1$ solution although an infinity of
minima occur up-stream). Furthermore, D$_{\sss N}$-solutions have the additional
feature that they may not exist for values of $\sigma$ lower than
$\sigma_{\sss N}(V_0)$. This was explained above in detail for the case $N=1$.
Figure \ref{table2} illustrates this generic behavior for a D$_2$-solution.

As for $N=1$, for this family there are generically two main types of
potential well, depending whether $n_{1,0}$ is larger (shallow potential) or
smaller (deep potential) than $n_{min}$ (left and right parts of Figure
\ref{table2}, respectively). In the case of a shallow potential the D$_2$
solution exists for any width below $\sigma_3(V_0)$; this is not the case for
deep potentials. We will not comment on Figure~\ref{table2} in great detail,
but we note that even in the simple case of a shallow potential interesting
bifurcations occur. Let us focus on this case. For simplicity we discard from
the discussion solutions forming cnoidal waves up-stream (i.e., for
$x\ge\sigma$). Then, for small widths, the only possible D$_2$ solutions are
even. For a certain width the minima of density occur exactly at $x=0$ and
$x=\sigma$ (it is easy to see that this width coincides with $\sigma_1^*(V_0)$
defined in Eq.~(\ref{asw6})). From there on, the previous even solutions still
exist, but new solutions appear. They correspond to a portion of a soliton
with its minimum before the well and one portion of cnoidal oscillation inside
the well (see Figure \ref{table2}, left part). This solution is degenerate in
the sense that there exists a symmetrical equivalent solution (where the
minimum of the soliton occurs beyond the well). It disappears when $\sigma =
\sigma_2(V_0)$. For $\sigma$ just below this value, one has exactly one
soliton out of the well and one cnoidal oscillation inside the well; hence the
trough of the solitonic part of the solution is sent to infinity (a feature
that is not clearly seen on Fig.~\ref{table2} due to numerical difficulties).
For larger values of $\sigma$ one has to oscillate more than once in the
region of the well, but these are D$_3$ and not D$_2$ solutions. On the other
hand, the even solutions still exist until $\sigma=\sigma_3(V_0)$. The
situation is slightly more complicated for deep potentials, but the basic
ingredients are the same as for shallow potentials and we present the
different allowed density profiles in Fig.~\ref{table2} (right part) without
detailed discussion.

\subsection{high beam velocity: $v_\infty / c_\infty >1$}\label{asw_v>c} 

\begin{figure}[thb]
\begin{center}
\includegraphics*[width=8cm]{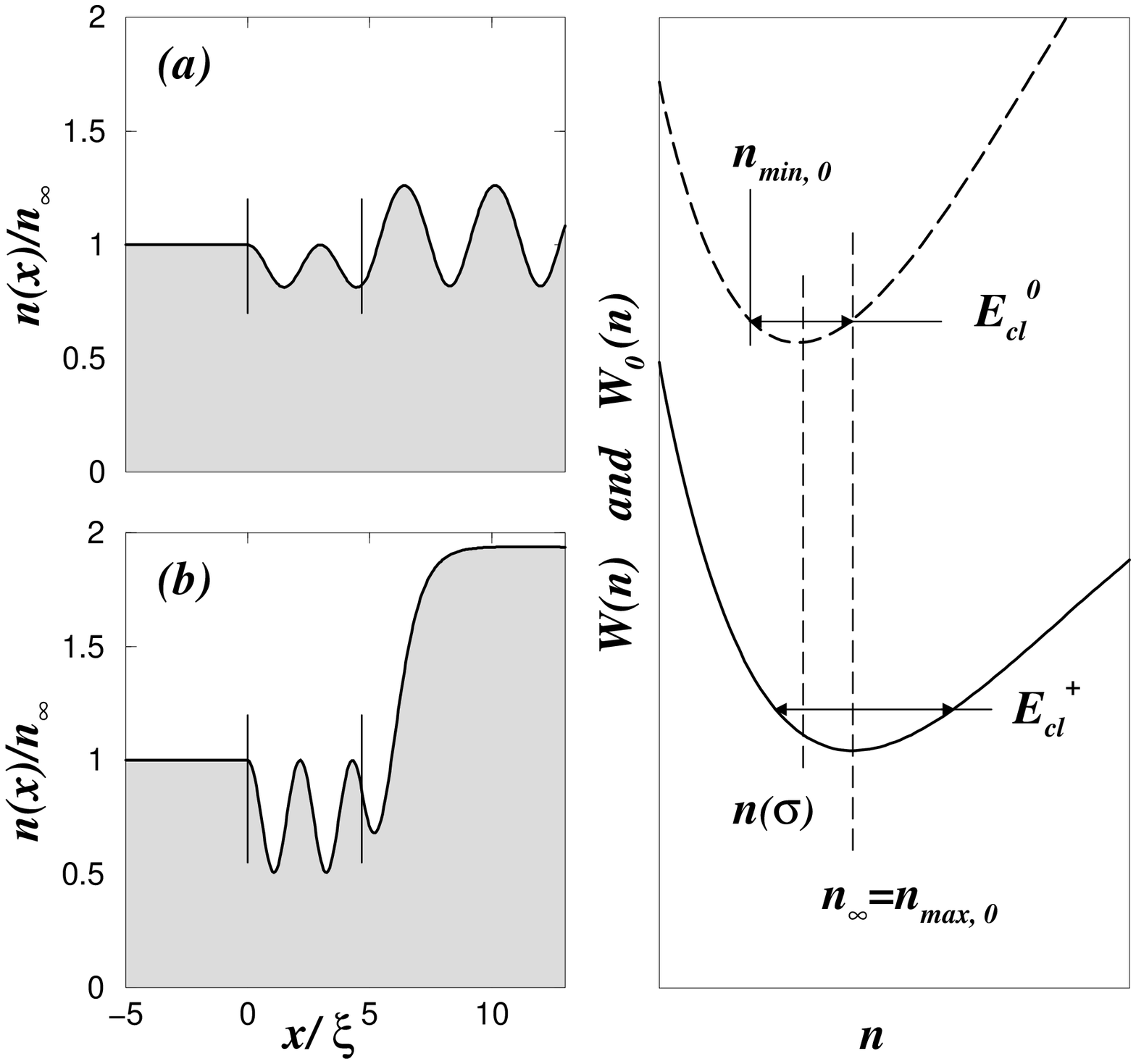}
\end{center}
\caption{\small A C-solution in an attractive square well. Left and right
parts as in Fig.~\ref{bump1}. Both density plots on the left part concern a
low density beam, with $v_\infty/c_\infty=1.6$ and a square well of width
$\sigma/\xi=4.68174$. Plot (a) is the generic case and corresponds to
$V_0\,\xi^2=0.1$. For these values of $V_0$ and $v_\infty$ the condition
(\ref{asw7}) is fulfilled for any $\sigma$. Plot (b) corresponds to a well of
depth $V_0\,\xi^2=0.5$. The parameters have been chosen in this case such that
$E_{cl}^+=W(n_2)$ and the C-solution is just about to disappear. The right
plot illustrates the behavior of the typical solution (corresponding to part
(a)) in the diagram $(n,W(n))$. The density is constant and equal to
$n_\infty$ for $x\le 0$ in $W(n)$. From $x=0$ to $x=\sigma$ the ``classical
particle'' evolves in $W_0(n)$ (with an energy $E_{cl}^0$). Finally the
up-stream density oscillates in the classical potential $W(n)$ as a cnoidal
wave.}
\label{Csolution}
\end{figure}

We now consider beam velocities $v_\infty$ larger (in absolute value) than
$c_\infty=c(n_\infty)$. In the language of Fig.~\ref{schema}, $n_\infty$ is in
this case of type $n_1$ (the minimum of the potential) and $E_{cl}^{-} =
W(n_1)$. The only possible flat solution far in the rear of the obstacle is a
flat and constant density, namely $n(x\le 0)=n_\infty$. The matching condition
(\ref{asw2}) yields $E_{cl}^+=E_{cl}^-+V_0(n_\infty-n(\sigma))$. One should
verify that $E_{cl}^+ \geq W(n_1=n_\infty)=E_{cl}^-$ (see Fig.~\ref{schema})
and this imposes $n(\sigma)\le n_\infty$. This inequality is trivially
satisfied, because the solution inside the well is a cnoidal wave with
$n_{max,0}=n_\infty$ (see Fig.~\ref{Csolution}). If we denote by $n_2$ (as in
Sec.~\ref{main}) the second zero of $dW/dn$ (the first one being
$n_1=n_\infty$) one has also to verify that $E_{cl}^+\le W(n_2)$. This imposes
\begin{equation}\label{asw7}
n(\sigma) \ge n_{\sigma,inf} 
= n_{\infty} -\frac{1}{V_0}\,[W(n_2)-W(n_1)] \; .
\end{equation}
 Once this condition is fulfilled, the up-stream solution is a cnoidal wave.
Because these solutions have a constant down-stream density, we denote them
C-solutions.

Two different cases are to be considered. The first and simpler one
corresponds to $n_{min,0}>n_{\sigma,inf}$. It occurs for high beam velocity
(when $W(n_2)-W(n_1)$ is large) and for shallow potentials (when $V_0$ is
small). In that case, the density of the cnoidal wave inside the well
oscillates between $n_{min,0}$ and $n_{max,0}$, and the matching at $x=\sigma$
is always possible. The incoming wave (for $x\ge\sigma$) is a cnoidal wave
corresponding to oscillations of the ``classical particle'' in the potential
$W(n)$. This behavior is illustrated in Fig.~\ref{Csolution}(a).

The other case corresponds to $n_{\sigma,inf}>n_{min,0}$. Then all the widths
are not acceptable, namely, $\sigma$ should be such that
$n(\sigma)>n_{\sigma,inf}$. If one denotes by $L_a$ the length the density
takes to go from the value $n_{\sigma,inf}$ to $n_{max,0}$ and by $L_0$ the
period of the cnoidal wave in the well, one has
\begin{equation}\label{asw8}
L_a=\frac{1}{\sqrt{2}}\, 
\int_{\surd n_{\sigma,inf}}^{\surd n_{max,0}} \frac{dA}{\sqrt{E_{cl}^0-W_0(A)}}
\qquad\mbox{and}\qquad
L_0=\sqrt{2}\,
\int_{\surd n_{min,0}}^{\surd n_{max,0}} \frac{dA}{\sqrt{E_{cl}^0-W_0(A)}}
\; .
\end{equation}
The allowed values of $\sigma$ are in the intervals
$[0,L_a]\cup[L_0-L_a,L_0+L_a] \cup[2L_0-L_a,2L_0+L_a]\cup...$.

\begin{figure}[thb]
\begin{center}
\includegraphics*[width=8cm]{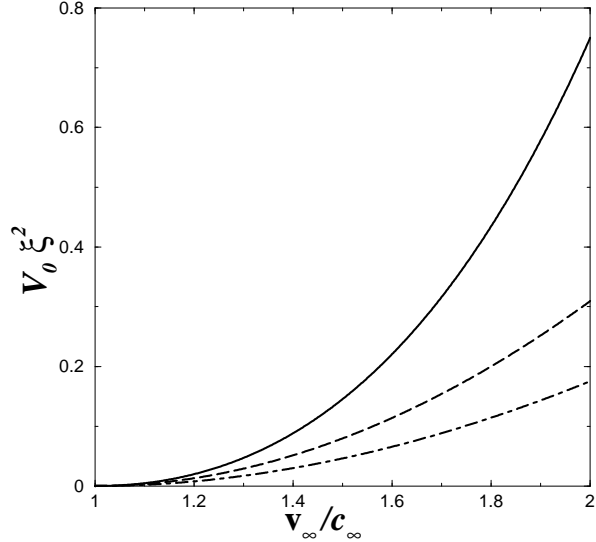}
\end{center}
\caption{\small Representation (for low density beams) of the part of the
plane $(v_\infty,V_0)$ where the C solution is allowed for any potential
width. The allowed region is located below the solid curve. We also display on
the same diagram the region where C solutions are allowed for any width of a
{\it repulsive} potential (see Sec.~\ref{rsw_v>c}). It corresponds to the
domain below the dashed curve for low density beams and below the dot-dashed
one for high density beams.}
\label{region}
\end{figure}

The region of validity of the first case $n_{min,0}>n_{\sigma,inf}$ can be
evaluated analytically in the low density regime. In this regime one obtains
$W(n_2)-W(n_1=n_\infty)=n_\infty\, F(v_\infty/c_\infty)\,/(4\,\xi^2)$ where
\begin{equation}\label{asw10}
F(z) =  
\left[
\frac{z^2}{4}\left(1+\sqrt{1+\frac{8}{z^2}}\right) - 1
\right]
\left[
\frac{5\,z^2}{4}+1-\frac{3\,z^2}{4}\sqrt{1+\frac{8}{z^2}}
\right]
\; .
\end{equation}
This yields $n_{\sigma,inf}/n_\infty = 1 - F(z) / (4\,V_0\,\xi^2)$ (we set
$z=v_\infty/c_\infty$). One also obtains $n_{min,0}/n_\infty=G(z,V_0\,\xi^2)$
where
\begin{equation}\label{asw11}
G(z,V_0\,\xi^2) =\frac{z^2+1}{2}+2\,V_0\,\xi^2-
\sqrt{\left(\frac{z^2+1}{2}+2\,V_0\,\xi^2\right)^2-z^2} \; .
\end{equation}
The region of validity of the condition $n_{\sigma,inf}<n_{min,0}$ (where all
the potential widths are acceptable) can be displayed in a diagram
$(z=v_\infty/c_\infty,V_0\,\xi^2)$. It corresponds to the region
\begin{equation}\label{asw11+}
1-\frac{F(z)}{4\,V_0\,\xi^2} < G(z,V_0\,\xi^2) \; ,\end{equation}
\noindent i.e., to the domain below the solid line in Fig.~\ref{region}.

\section{The repulsive square well}\label{rsw}

In this section we consider a simple repulsive potential, namely, we take
$V_\parallel(x)$ to be zero everywhere except in a finite region $0\le
x\le\sigma$ where it takes the constant value $V_0$ ($V_0>0$). This type of
potential cannot correspond to a bend in the guide, but it can be realized
with a (blue detuned) far off-resonant laser field.

Eq.~(\ref{asw1}) still holds after changing the sign of $V_0$.
Eq.~(\ref{asw2}) holds also, but we rewrite it here with the appropriate sign
for future reference,
\begin{equation}\label{rsw1}
E_{cl}^- - V_0\,n(0) = E_{cl}^0 = E_{cl}^+ - V_0\,n(\sigma) \; .
\end{equation}
\noindent and similarly one has here (instead of (\ref{w0at})) 
\begin{equation}\label{w0rep}
W_0(n)=W(n)-V_0\,n
\end{equation}
The potential $W_0(n)$ has, for $V_0$ relatively small, the same behavior as
$W(n)$, namely one minimum at $n_{1,0}$ and one maximum at $n_{2,0}$, with
$n_1 < n_{1,0} < n_0 < n_{2,0} < n_2$ (the notations are defined in
Sec.~\ref{main}, see Fig.~\ref{schema}). On the other hand, for large $V_0$,
$W_0$ is a monotonically decreasing function of $n$. The transition between
the two regimes occurs when $\mu-V_0=\epsilon(n_0)+\Phi^2/(2\,n_0^2)$. Hence,
the terms ``low'' or ``large'' $V_0$ we just defined (we also speak below of
``weak'' and ``strong'' potentials) are not intrinsic properties of the
potential, but depend on the chemical potential $\mu$ and on the flux $\Phi$
of the incoming beam (this remark is made quantitative in Sec.~\ref{rsw_v>c}
below).

Amongst all the possible solutions, we choose again to select those
corresponding to a flat density at left infinity, ($n(x)\to n_\infty$ as $x\to
-\infty$), with a negative velocity ($v(x)\to v_\infty<0$). Under these
conditions, the transmission modes through repulsive potentials appear to be
simpler than for the attractive ones discussed in Sec.~\ref{asw}, as we are
now going to show.

\subsection{high beam velocity: $v_\infty / c_\infty >1$}\label{rsw_v>c}

The stationary solutions of a supersonic beam encountering a repulsive square
well have important similarities with the C-solutions of Sec.~\ref{asw} and
will be given the same name. For weak potentials ($V_0 < \mu-\epsilon(n_0) +
\Phi^2/(2\,n_0^2)$) the solutions exist whatever the value of $\sigma$, and
their shape is very similar to the transmission mode illustrated in
Fig.~\ref{Csolution}(a). The main difference is that here, one has $n(x)\ge
n(0)=n_\infty$ in the region of the well (whereas the reverse inequality holds
for C-solutions in an attractive well).

\begin{figure}[thb]
\begin{center}
\includegraphics*[width=8cm]{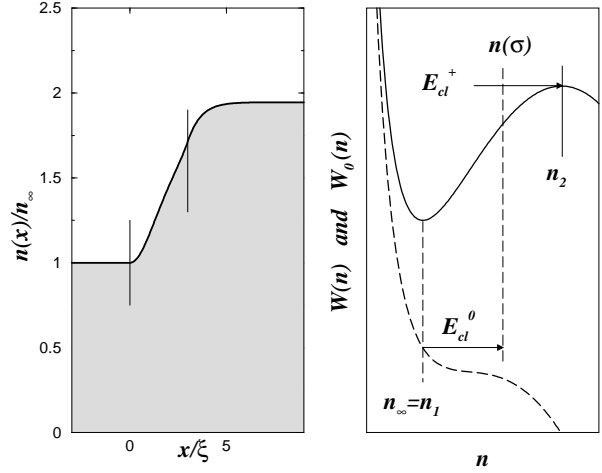}
\end{center}
\caption{\small A C-solution in a repulsive square well. The parameters have
been chosen in this case such that $E_{cl}^+=W(n_2)$ and the C-solution is
just about to disappear. Left and right parts as in Fig.~\ref{bump1}. The left
plot concerns a low density beam, with $v_\infty/c_\infty \simeq 1.6$ and a
square well of width $\sigma/\xi=3$ and depth $V_0\,\xi^2=0.1$. The right plot
illustrates the behavior of this solution in the diagram $(n,W(n))$. The
down-stream density is constant and equal to $n_\infty$. From $x=0$
to $x=\sigma$ the ``classical particle'' evolves in $W_0(n)$ (with an energy
$E_{cl}^0$), then it evolves back in $W(n)$, just on the separatrix.}
\label{cstar}
\end{figure}

On the other hand, for strong potentials the solution in the region of the
well is not (as for weak potentials) a cnoidal wave. In this region the
density increases monotonically from $x=0$ up to $x=\sigma$; then
``classical'' motion occurs in $W(n)$ and the up-stream density profile
further oscillates as a cnoidal wave. This means that there exists a maximum
value of $\sigma$ above which no solution can be found in a strong potential.
When $\sigma$ reaches this maximum value, the solution for $x\ge\sigma$ is a
portion of a soliton and the whole solution has a step-like shape going from
$n_\infty = n_1$ (far in the rear of the obstacle) up to $n_2$ at
$x\rightarrow +\infty$. The behavior in this limiting case is illustrated in
Fig.~\ref{cstar}. There is here a difference with the case of a strong
attractive square well: when $\sigma$ has exceeded this maximum value, no
other stationary solution appears for larger widths. The reason is that in an
attractive square well, the solution in the region of the well is periodic,
whereas it increases here without limit for large wells.

Due to the similarities with the case of an attractive potential, it is
interesting here also to determine more precisely for which beams a stationary
supersonic solution can be found (for weak potentials) for all values of
$\sigma$. An analytical treatment is here possible in both the low and high
density regimes. The two cases can be treated on the same footing by
introducing an index $\nu$ with $\nu=1$ in the dilute regime and $\nu=1/2$ for
high densities. One has
\begin{equation}\label{rsw3}
\epsilon(n)=\epsilon_0+
\frac{1}{2\,\xi^2} \,\left(\frac{n}{n_\infty}\right)^\nu 
\qquad\mbox{and}\qquad
v^2=\frac{\nu\,z^2}{2\,\xi^2} \,\left(\frac{n_\infty}{n}\right)^2 \; ,
\end{equation}
\noindent where $z=v_\infty/c_\infty$. One then obtains $n_0/n_\infty =
z^{\frac{2}{\nu+2}}$ ($n_0$ is defined in Sec.~\ref{main}, Fig.~\ref{schema}).
The condition for a repulsive potential to be considered as weak is thus
$\mu-V_0 \ge \epsilon(n_0) + \Phi^2/(2n_0^2) = \epsilon_0+ (n_0/n_\infty)^\nu
\, (1+\frac{\nu}{2}) / (2\,\xi^2)$. Since
$\mu=v_\infty^2/2+\epsilon(n_\infty)$, this reads
\begin{equation}\label{rsw4}
V_0\,\xi^2\le \frac{1}{2}+\frac{\nu\,z^2}{4}
-\frac{1}{2}\left(1+\frac{\nu}{2}\right)\,z^{\frac{2\,\nu}{\nu+2}} \; .
\end{equation}
This region of weak potential corresponds in Fig.~\ref{region} to the domain
below the dashed curve in the dilute regime $\nu=1$, and below the dot-dashed
curve in the high density regime $\nu=1/2$. As stated in the beginning of this
Section, the terms weak or strong potential do not characterize an intrinsic
property of the well. At $v_\infty=c_\infty$ for instance, all the potentials
are ``strong''.

	In view of future experimental studies of the system, it is also interesting
to determine (in the regime of ``strong potentials'') the critical value of
$\sigma$ after which no C-solution can be found. We denote this value by
$L_{r}$. Following a reasoning similar to the one of Sec. \ref{asw_v>c} one
obtains
\begin{equation}\label{rsw5}
L_r=\frac{1}{\sqrt{2}}\, 
\int^{\surd n_{\sigma,sup}}_{\surd n_{min,0}}
\frac{dA}{\sqrt{E_{cl}^0-W_0(A)}}
\qquad\mbox{where}\qquad
 n_{\sigma,sup} = n_{\infty} +\frac{1}{V_0}\,[W(n_2)-W(n_1)]
\; .
\end{equation}
	In the low density regime $L_r$ can be expressed in terms of an elliptic
integral:
\begin{equation}\label{rsw6}
\frac{L_r}{\xi} = \int_0^{F(z)/(4V_0\xi^2)}
\frac{dx}
{\sqrt{2\,x}\left[x^2+x(4\,V_0\,\xi^2+1-z^2)+4\,V_0\,\xi^2\right]^{1/2}}
\simeq
\frac{1}{V_0\,\xi^2}\,\left(\frac{F(z)}{8}\right)^{1/2}\; .
\end{equation}
	Note of course, that Eq.~(\ref{rsw6}) is meaningful only when the condition
(\ref{rsw4}) is violated. $L_a$ in Eq.~(\ref{asw8}) can also be defined with a
similar expression in the low density regime. Both expressions are well
approximated by the left part of (\ref{rsw6}), meaning that one has $8 \,
(L_{r/a}\,V_0\,\xi)^2\simeq F(z)$. We will see below that this corresponds to
approximating the potential by a $\delta$-function (cf. Sec.~\ref{simple},
Eq.~(\ref{delta3})).

\subsection{low beam velocity: $v_\infty / c_\infty <1$} 

It is easy to check that in this case, only D$_1$ solutions can be observed
(there are no other D-solutions, and B-solutions are forbidden). The
down-stream solution starts at left infinity from a density $n_\infty$ which
is of type $n_2$ in the terminology of Sec.~\ref{main}. $n(x)$ decreases
from this value, and in the diagram $(n,W(n))$ the fictitious classical
particle evolves in the potential $W_0(n)$ during a ``time'' $\sigma$, and
then evolves in $W(n)$ again. Since one should verify that $E_{cl}^+ \leq
W(n_2)=E_{cl}^-$, from (\ref{rsw1}) this imposes $n(\sigma)\le n(0)$. If the
inequality is strict, the solution for $x>\sigma$ is a cnoidal wave. If
$n(\sigma)= n(0)$ the final solution is a portion of a soliton.

\begin{figure}[thb]
\begin{center}
\includegraphics*[width=8cm]{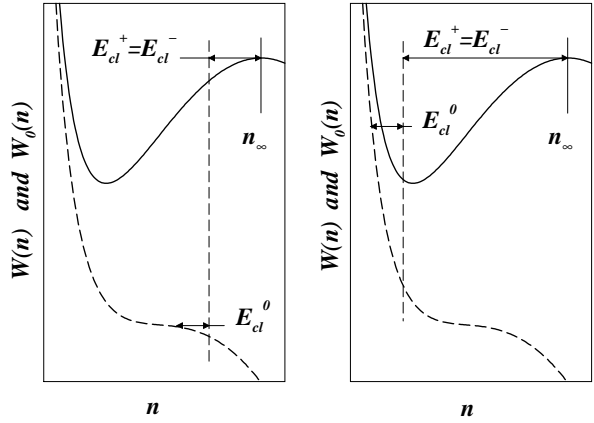}
\end{center}
\caption{\small Schematic representation in the $(n,W(n))$ diagram of the
behavior of D$_1$-solutions in a strong repulsive square well
($v_\infty/c_\infty <1$). The solid curve represents $W(n)$ and the dashed one
$W_0(n)$. For this type of potential there are two possible solutions for a
given value of the width $\sigma$ (as discussed in the text). This is
illustrated by the figure, namely the ``time'' the fictitious particle spends
in the potential $W_0$ is the same in the left and right plots. For simplicity,
the examples we give here are drawn in the particular case $n(0)=n(\sigma)$.}
\label{repuls}
\end{figure}

Let us consider a strong potential first. If the well is narrow ($\sigma \to
0$), there are two possible solutions, depending whether, in the diagram
$(n,W(n))$, one ``jumps'' rapidly or not from $W$ to $W_0$. This is
illustrated on Fig.~\ref{repuls}. When $\sigma$ increases, these two solutions
merge and disappear. This behavior was already observed by Hakim in the case
of a model repulsive $\delta$-potential \cite{Hak97}. Hakim showed (for the one
dimensional Gross-Pitaevskii equation, i.e., in the dilute regime) that the
solution which ``jumps late'' to $W_0$ (right part of Fig.~\ref{repuls}) is
unstable, and argued convincingly that the other is stable. The same situation
is expected to occur here.

In the case of a weak potential there are also two types of solution, but
they do not disappear when $\sigma$ increases. The point is that one can have
here $E_{cl}^0$ arbitrarily close to the separatrix energy $W_0(n_{2,0})$, and
the period of motion in the potential $W_0$ can thus be made as large as
desired. From (\ref{rsw1}), one has $E_{cl}^0=W_0(n_{2,0})$ if $n(0)=n_0^*$
with
\begin{equation}\label{rsw2}
n_0^*=n_{2,0}+\frac{1}{V_0}\left[W(n_2)-W(n_{2,0})\right] \; .
\end{equation}
The expression ``jumping soon'' (or ``late'') from $W$ to $W_0$ that we used
in the discussion of strong potentials refers, for a weak potential, to the
case where $n(0)>n_0^*$ (or $n(0)<n_0^*$). If $n(0)>n_0^*$, the density
remains larger than $n_{2,0}$ and reaches this value at its minimum in the
limit $\sigma\to\infty$. If $n(0)<n_0^*$, $E_{cl}^0>W_0(n_{2,0})$ and the
fictitious classical particle evolves above the separatrix in the potential
$W_0$. As in the case of a strong potential, this solution is expected to be
unstable. The other one is certainly stable since one can show that it is
identical to the result of perturbation theory in the limit of a very weak
potential.

\section{Simple solutions in the presence of an obstacle}\label{simple}

The aim of this section is to study, by means of perturbation theory, some
simple solutions of Eq.~(\ref{eq1}) valid for a generic potential
$V_\parallel(x)$. We will argue that near the speed of sound this approach
fails, and that in this regime any potential can be approximated by a
$\delta$-peak. We will then study the scattering modes of the condensate in
the presence of this potential. It allows for a qualitative and simple
understanding of the solutions obtained for more realistic potentials in the
previous sections. Some of the results presented here have already been
obtained by Hakim \cite{Hak97}, who considered repulsive potentials only, in a
slightly less general setting.

We again restrict the analysis to those transmission modes tending to a flat
density at $x \to -\infty$, $n(x \to -\infty) \to n_\infty$, with a negative
velocity $v_\infty$. These are of the form $A(x)=A_\infty+\delta A(x)$ (with
$A^2_\infty=n_\infty$). Denoting by $c_\infty$ the sound velocity at density
$n_\infty$ (Eq.~(\ref{sound})), a perturbative treatment of Eq.~(\ref{eq1})
yields
\begin{equation}\label{s1} \delta
A^{\prime\prime} + 4\,(v_\infty^2-c_\infty^2)\, \delta A = 
2\, A_\infty\, V_\parallel(x) \; . 
\end{equation} 
The solutions of (\ref{s1}) that tend to zero when $x\to-\infty$ are of the
form
\begin{equation}\label{s2} 
\delta A (x) = \left\{ \begin{array}{lcc}
-\frac{A_\infty}{k} \int_{-\infty}^{+\infty} V_{\parallel}(y) \exp\{-k
|x-y|\}\, dy & \mbox{when} & v_\infty/c_\infty <1 \; , \\ & & \\
\frac{2\, A_\infty}{k} \int_{-\infty}^{x} V_{\parallel}(y)
 \sin\{k (x-y)\}\, dy &
\mbox{when} & v_\infty /c_\infty >1 \; , \end{array}\right.
\end{equation}
where $k=2 \, |v^2_\infty-c^2_\infty|^{1/2}$.

Denoting by $\sigma$ the typical range of the potential $V_\parallel$, in the
limit $k\sigma\gg 1$ the Green function of (\ref{s1}) is almost a delta peak
and Eqs.~(\ref{s2}) take the simple form
\begin{equation}\label{s2+}
\delta A (x) = \mbox{sgn}\, (v_\infty /c_\infty - 1) \; \frac{2\,
A_\infty}{k^2} \; V_\parallel(x) \; .\end{equation}

This result may seem unnatural at large velocities. Indeed, for a repulsive
potential for instance, the density {\it increases} in the region of the
potential. This kind of behavior was already found in Ref.~\cite{Law00}, where
a very special potential was designed for which this phenomenon occurs at any
$v_\infty /c_\infty >1$. From (\ref{s2+}) we see that similar motional dressed
states exist for any potential in the limit of very large velocities (when
$k\sigma\gg 1$).

In the case $v_\infty /c_\infty >1$, an asymptotic evaluation of (\ref{s2})
yields far ahead of the obstacle (in the limit $k\,x\gg 1$)
an amplitude of the form
\begin{equation}\label{s2++}
\delta A (x) = \frac{2\, A_\infty}{k}\; 
\mbox{Im}\,\left\{e^{i\,k\,x} \hat{V}_\parallel (k) \right\} 
+ \frac{2\,A_\infty}{k^2} \; V_\parallel(x) \; +
{\cal O} \left(\frac{1}{k^3}\right) \; ,
\end{equation}
where $\hat{V}_\parallel (k) = \int_{-\infty}^{+\infty} dx \exp(-ikx)
V_\parallel(x)$ is the Fourier transform of $V_\parallel(x)$. This shows that,
for $v_\infty /c_\infty >1$, Eq.~(\ref{s2}) corresponds to a C-solution. The
fact that the up-stream solution oscillates with a wave-vector $k$ corresponds
to the stationarity condition $v_p(k)=-v_\infty$ (see the end of
Sec.~\ref{main}). The wake is characterized by a wavelength $2\pi/k$ that
depends on the velocity of the beam (it decreases when $|v_\infty|$
increases). This is due to the particular form of the dispersion relation
(\ref{eq3}), and does not occur above the Landau critical velocity in liquid
helium for instance, where the location of the roton minimum fixes the
wavelength of the wake \cite{Pit84}.

When the beam velocity is lower than the speed of sound, for attractive (resp.
repulsive) potentials Eq.~({\ref{s2}) describes a B$_1$ (resp. D$_1$)
solution. In the attractive case for instance, the bump density measured with
respect to the constant density $n_\infty$ contains a number $\Delta N$ of
atoms given by $\Delta N = \int_{-\infty}^{+\infty} \delta n(x) \, dx
\approx  -(4n_\infty/k^2) \int_{-\infty}^{+\infty} dx\, V_\parallel(x)$.
This formula diverges when $v_\infty / c_\infty \to 1$ since $k \to 0$. In that
limit, however, the perturbative treatment is not justified. Indeed,
Eq.~(\ref{s2}) gives a sensible result only if $|\delta A|\ll A_\infty$.
Denoting by $V_0$ the typical value of the potential $V_\parallel$, this
reads
\begin{equation}\label{s3}
\left\{ \begin{array}{lcc}
V_0 \ \sigma /k \ll 1 & \mbox{when} 
& k \sigma \ll 1 \; , \\ & & \\
V_0 /k^2 \ll 1 & \mbox{when} & k \sigma \gg 1 \; . \end{array}\right.
\end{equation}
These conditions are satisfied only if the beam velocity $v_\infty$ and the
sound velocity $c_\infty$ are not too close. 

From Eqs.~(\ref{s2}) we see that for $v_\infty / c_\infty <1$ the typical
length scale of variation of $\delta A$ is proportional to $k^{-1}$. If
$|v_\infty|$ approaches $|c_\infty|$ from below, this length scale diverges
and the spatial extension of $\delta A$ increases indefinitely. In this case
it is legitimate to approximate $V_\parallel$ by a $\delta$-function. Hence, we
do not pursue the perturbative treatment any longer and turn now to the study
of solutions in a $\delta$-peak potential.

%

Consider a potential of the form $V_\parallel(x)=\lambda\,\delta(x)$ with
$\lambda$ positive or negative. A realistic $V_\parallel$ can be approximated
by such a potential if its typical length scale $\sigma$ is much lower than
the healing length $\xi$. In this case the approximation is valid for any beam
velocity. As discussed above, any potential $V_\parallel$ can be approximated
by a $\delta$-function when $v_\infty$ approaches $c_\infty$, i.e., in the
limit $k\sigma\ll 1$.

The allowed transmission modes of Eq.~(\ref{eq1}) when $V_\parallel$ is a
$\delta$-potential are very simple. They are obtained by joining together the
solutions of two straight guides (see Sec. \ref{main}), one down-stream
($x<0$) and the other up-stream ($x>0$), with the matching condition
\begin{equation}\label{delta1}
A'(0^+)-A'(0^-)=2\,\lambda\,A(0) \; .
\end{equation}
 The integration constant $E_{cl}$ of Eq.~(\ref{eq2}) changes discontinuously
from the value $E_{cl}^-$ for $x<0$ to the value $E_{cl}^+$ for $x>0$
($E_{cl}^{\pm}=\frac{1}{2}[A'(0^\pm)]^2+W(n(0))$). Using (\ref{delta1}) these
two values are related through $E_{cl}^+= E_{cl}^- + \lambda\,A(0)
\,[A'(0^-)+A'(0^+)]$.

\begin{figure}[thb]
\begin{center}
\includegraphics*[width=7cm]{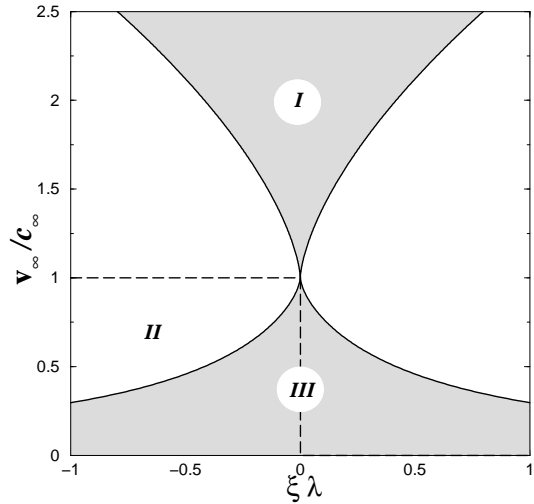}
\end{center}
\caption{\small Domains of existence of the different solutions occurring for
a dilute beam in a potential $V_\parallel(x) = \lambda\,\delta(x)$. Region I
occurs only for $v_\infty /c_\infty >1$ and corresponds to C-solutions. Region
II corresponds to B$_1$-solutions and occupies the domain under the dashed
line (it occurs only for attractive potentials and for $v_\infty /c_\infty
<1$). Region III corresponds to D-solutions: D$_1$-solutions for $\lambda>0$
and D$_2$ solutions when $\lambda<0$.}
\label{cas}
\end{figure}

Let us first consider the case $v_\infty /c_\infty >1$. We know from Secs.
\ref{asw_v>c} and \ref{rsw_v>c} that one should observe C-solutions. One has
in this case a constant down-stream density $n_\infty = A_\infty^2$
and $E_{cl}^{-}=W(n_\infty)$. The matching condition imposes
$E_{cl}^{+} = E_{cl}^{-} + 2\,\lambda^2\,n_{\infty}$ and $A'(0^+) = 2 \lambda
A_\infty$. In order to have a finite solution at $x=+\infty$ one should
moreover satisfy $W(n_\infty)\leq E_{cl}^+ \leq W(n_2)$. This fixes an upper
bound for the intensity of the perturbation, given by $2 \, \lambda^2 n_\infty
\leq W(n_2)-W(n_\infty)$. This is the criterion for observing C-solutions. In
the low-density limit this relation takes the analytical form
\begin{equation}\label{delta3}
8\, \lambda^2\,\xi^2 \leq F\left(\frac{v_\infty}{c_\infty}\right)\; ,
\end{equation}
where the function $F$ is defined in (\ref{asw10}). It corresponds to the
region I of Fig.~\ref{cas}.

Let us now consider the case $v_\infty /c_\infty <1$. The first and simpler
solution is the B$_1$-solution which occurs for attractive potentials. The
corresponding matching condition (\ref{delta1}) reads in this case
$n'(0)/n(0)=-2\,\lambda$. It can be fulfilled whatever the value of $\lambda$
($\lambda<0$) because the function $(n'/n)^2=8\,n^{-1}[E_{cl}-W(n)]$ can be
made arbitrarily large (it grows as $8\,\varepsilon(n)/n$ for large $n$).
The domain of existence of B$_1$-solutions corresponds to region II in
Fig.~\ref{cas}.


D-solutions can also be observed if $v_\infty /c_\infty <1$. For $\lambda>0$
these are the D$_1$ solutions whose behavior in the $(n,W(n))$ diagram is
illustrated in Fig.~\ref{repuls} (in the case of a depth of finite width
though). For $\lambda<0$ these are D$_2$-solutions such as those shown in
Fig.~\ref{table2} for $\sigma\to 0$. The limiting case (beyond which these
solutions disappear) is obtained for two symmetric portions of solitary waves.
The matching condition (\ref{delta1}) then fixes the maximum value of
$|\lambda|$, which corresponds in both cases to
\begin{equation}\label{delta4}
|\lambda|\,\xi = 
K\left(\frac{v_\infty}{c_\infty}\right)
\qquad\mbox{where}\qquad
K(z)=
\frac{1}{4\,z} \,
\left\{-8\,z^4-20\,z^2+1+(1+8\,z^2)^{3/2}\right\}^{1/2}\; .
\end{equation} 
The corresponding domain is shown in Fig.~\ref{cas} (region III).

\section{Experimental considerations}\label{experiments}

	In view of future experimental observation of the different flows past an
obstacle, we now evaluate the orders of magnitude of the relevant
dimensionless parameters identified above. For concreteness we consider a beam
such as the one in preparation at the ENS \cite{dgo}: $^{87}$Rb atoms are
guided along the $x$ direction, with a harmonic transverse confinement
($\omega_\perp=2\,\pi\times 500$ Hz and $a_\perp=0.5\; \mu$m). The beam has a
velocity $v_\infty$ of order of 0.5 m/s and a flux $\Phi$ varying from $10^4$
to $10^8$ atom.s$^{-1}$. Hence the quantity $n_\infty\,a_{sc}$ varies from
$10^{-4}$ to 1. For a rough estimate of the order of magnitude of the relevant
parameters, we will consider that this corresponds to the low density limit
$n_\infty\,a_{sc}\ll 1$. Then, the healing length (defined in (\ref{adia7}))
and the speed of sound vary from $\xi=25\,\mu$m and $c_\infty =
20\,\mu$m.s$^{-1}$ (for $n_\infty\,a_{sc}=10^{-4}$) to $\xi = 0.25\,\mu$m and
$c_\infty = 2$ mm.s$^{-1}$ ($n_\infty\,a_{sc}=1$). Note that the beam velocity
is much larger than the speed of sound (the quantity $v_\infty/c_\infty$ is of
order $10^3$ at least).

	If the obstacle is a bend of constant radius of curvature $R_c$ and opening
angle $\Theta$, $V_0=-1/(8\,R_c^2)$ and $\sigma=\Theta\,R_c$. A reasonable
order of magnitude is $R_c=5\,a_\perp$, leading to $V_0\,\xi^2= \frac{1}{200}
(\xi/a_\perp)^2$ which varies roughly from $10^{-3}$ to 10. For $\Theta=\pi/2$
one has $\sigma=8\,a_\perp$ and the obstacle can safely be treated
perturbatively because in this case $k \sigma \sim 10^4 \gg 1$ and
Eq.~(\ref{s3}) is satisfied since 
$$
V_0 / k^2 \approx 10^{-8} \ll 1 \ . 
$$
For this configuration one thus expects profiles in agreement with
Eq.~(\ref{s2}) (case $v_\infty /c_\infty >1$), i.e. C-solutions. Also,
since we are in the regime $k\,\sigma\gg 1$ Eq.~(\ref{s2+}) holds, meaning
that the wake ahead of the obstacle is very weak, and that there is a decrease
in the density in the region of the bend. However, this decrease is extremely
small: it corresponds to a number of atoms smaller than unity, and under these
conditions nothing noticeable is expected to occur in the bend.

The situation changes drastically if the obstacle is due to a transverse laser
beam because the potential can be made much stronger. It can moreover be
attractive or repulsive depending on the laser's frequency, and also the
velocity of the obstacle relative to the beam can be modified by using an
acousto-optic deflector.

We consider a laser with power $P=70$ mW and a wave length $\lambda_{\sss L}$
varying from 780 to 790 nm (the atomic transition corresponds to a wavelength
$\lambda_{\sss A}=780.2$ nm and has a natural width $\Gamma=12\,\pi$ MHz). The
laser beam has a typical waist $\sigma$ of the order of 50 $\mu$m. Then one
obtains for the transverse potential
\begin{equation}\label{exp1}
V_\parallel(x) = V_0\,\exp\{- \frac{2\, x^2}{\sigma^2} \} 
\qquad\mbox{where}\qquad 
V_0 = \frac{\hbar\,\Gamma^2\,P}{4\,\pi\,\delta\,\sigma^2\,I_{\sss S}} \; ,
\end{equation}
\noindent $\delta=\omega_{\sss A}-\omega_{\sss L}$ being the detuning and
$I_{\sss S}=16$ W.m$^{-2}$ is the saturation intensity. This yields
\begin{equation}\label{exp2}
V_0\,\sigma^2 = \frac{1}{4\,\pi}\,
\left(\frac{\Gamma^2}{\delta\,\omega_\perp}\right) \, 
\left(\frac{P}{I_{\sss S}\,a_\perp^2}\right) \sim \pm 10^8 \; .
\end{equation}
 In this case one is in the regime $k\,\sigma\gg 1$, with $V_0/k^2\sim
10^{-2}$ or $10^{-1}$. Hence one is on the edge of applicability of the
perturbative approach. By changing the wavelength of the laser or its velocity
with respect to the beam, one may enter into the non-perturbative regime. One
can still find C-solutions for this type of potential if the parameters of the
system remain in the appropriate region of figure~\ref{region}. C-solutions
exist for any value of $\sigma$ if $V_0\,\xi^2\le
\frac{1}{16}\,(v_\infty/c_\infty)^4$ in the case of an attractive potential,
and if $V_0\,\xi^2\le \frac{1}{4}\,(v_\infty/c_\infty)^2$ in the case of
repulsive potential (since $v_\infty/c_\infty\gg 1$ we consider here the
asymptotic versions of Eqs.~(\ref{asw11+}) and (\ref{rsw4})). By modifying the
value of the healing length and of the beam velocity these conditions can
easily be satisfied or violated (especially in the case of repulsion). In the
region where these conditions are violated one can experimentally study the
transition from a stationary flow (of type C) to a time dependent one. For
instance, it would be of great interest to study the modification of the drag
at the transition.

Just at the boundary between the two regions, for a repulsive potential the
density profile has the behavior shown in Fig.~\ref{cstar} (for an attractive
potential the density profile corresponds to Fig.~\ref{Csolution}(b)). The
density has a step-like shape, and the beam ahead of the obstacle has a
velocity lower than the speed of sound. One has just at the transition
$\sigma=L_r$ (or $L_a$ in the case of an attractive potential) with
$8\,(V_0\,L_r\, \xi)^2\simeq F(v_\infty/c_\infty)\simeq
(v_\infty/c_\infty)^4/4$ (see the end of Sec.~\ref{rsw_v>c}). From the
estimate (\ref{exp2}) this occurs for $\xi/\sigma\sim 10^{-2}$ or $ 10^{-3}$.
For a laser, one can tune the waist $\sigma$ by a factor of order 5 say, and
more important, by changing the density one can modify the value of $\xi$ and
indeed reach the appropriate regime. This would have a very important effect
on the beam, since the velocity ahead of the obstacle would be {\it lower}
than the speed of sound whereas it is of the order of 1 m.s$^{-1}$ down-stream.
Accordingly, the density along the beam would go from $n_\infty=n_2$ to $n_1$
(see Fig.~\ref{cstar}). It is not difficult to see that, in the dilute regime,
in the limit $v_\infty /c_\infty \gg 1$ one has $n_2\simeq n_1
(v_\infty/c_\infty)^2/4$. Thus the down-stream beam density is divided by a
factor of order $10^{6}$ with respect to the up-stream one and, by
conservation of the flux, the velocity is multiplied by the same factor. The
beam velocity ahead of the obstacle is then of the order of a few micron per
second!

\section{Conclusions}\label{conclusion}

In this paper we have studied the different stationary profiles of a
Bose-Einstein condensed beam propagating through a guide with an obstacle. The
beam far down-stream is characterized by its velocity $v_\infty$ and density
$n_\infty$ (or equivalently, its healing length $\xi$). The obstacle is
represented by a one dimensional square well of depth $\pm V_0$ and width
$\sigma$. Let us denote by $\Delta$ the region along the guide where the
potential is different from zero. Our study allows to identify the relevant
dimensionless parameters governing the flow across the obstacle. These are
$n_\infty\,a_{sc}$, $v_\infty/c_\infty$, $\sigma/\xi$ and $V_0\,\xi^2$.
By varying these parameters one obtains a wide range of density profiles (we
identify three main stationary families denoted as B, C and D-solutions). We
have numerically checked \cite{nous} that similar results are obtained for
potentials other than square wells (such as $V_\parallel(x)=\pm
V_0\exp\{-x^2/\sigma^2\}$ for instance) and we thus believe that our analysis
of the flow is quite general.

The richest case occurs when the external potential is attractive. In the
subsonic regime (beam velocity lower than the corresponding speed of sound)
the simplest solution is a symmetric density having a peak in $\Delta$
(B-family). These solutions may have density oscillations in $\Delta$, and are
always symmetric. Another type of transmission mode is a soliton-like
depressed solution pinned to the obstacle (D-family), which may also have
density oscillations in $\Delta$ and, unlike the B-family, a wake
up-stream. Finally, the supersonic transmission modes (C-solutions) possess,
in the simplest case, a density trough in $\Delta$ and are constant outside.
They may also have density oscillations in $\Delta$ and an up-stream wake.

For a repulsive potential, in the subsonic regime the transmission modes are
of type D with no density oscillations in $\Delta$. In the supersonic case,
the modes are of type C but with a density peak instead of a trough. Step-like
solutions of increasing density across the obstacle (with or without wake)
also exist. Specifically, we have identified an interesting (and
experimentally reachable) regime where the beam is almost stopped by a
repulsive obstacle and gains ahead of it several orders of magnitude
in density (see Sec. \ref{experiments}).

An important aspect of the problem that remains open are considerations
related to the stability of the solutions. Some related work is in progress
\cite{nous}. We just note here that the limiting C-solution shown in
Figs.~\ref{Csolution}b and \ref{cstar} can be turned into a D-solution by
exchanging the down-stream and the up-stream behavior of the flow. However,
these are part of a continuous family of flow patterns and are probably
instable. The C-solutions on the other hand, as selected by the radiation
condition (see the end of Sec.~\ref{main}), are the only acceptable ones for
$v_\infty/c_\infty>1$. Another aspect concerns Bose gases with attractive
intra-atomic interactions, which may also be treated with our formalism. In
this case the potential $W(n)$ introduced in Sec.~\ref{main} has a single-well
shape, and the number of different transmission modes whose density tends to a
constant at the input of the guide is greatly reduced with respect to the
case of repulsive interactions considered here.

We conclude by noting that a branch of BEC that is now expanding is the
non-linear counterpart of transport experiments of mesoscopic physics in
condensed matter. In the latter case the coherent transport of two-dimensional
electron gases through various geometries has been considered in great detail.
Future developments of transport experiments of Bose condensates should
extend those to non-linear regimes.\\

\noindent {\large \bf Acknowledgments}

\bigskip \noindent It is a pleasure to thank S. Brazovskii, C. Schmit, J.
Treiner and D. Ullmo for fruitful discussions. Special thanks to F. Dalfovo
and S. Stringari for discussing the results of this paper and bringing
Ref.~\cite{Pit84} to our attention. We also would like to thank warmly D.
Gu\'ery-Odelin for numerous fruitful exchanges.

\appendix
\section{}\label{goldstone}
\setcounter{equation}{0}

In this Appendix we recall the conditions for existence of a bound state in
a bend of an ordinary wave guide (i.e. without non-linear terms in the
Schr\"odinger equation) and put the adiabatic limit used in the text on
firmer mathematical grounds.

Let us consider a curve ${\cal C}$, of parametric equation $\vec{r}_{\sss C}
(x)$, $x$ being a curvilinear abscissa along ${\cal C}$. The Frenet frame
$(\vec{t},\vec{n},\vec{b}\,)$, the curvature $\kappa(x)$ and the torsion
$\tau(x)$ are defined by $\vec{t}=d\vec{r}_{\sss C}/dx$,
$d\vec{t}/dx=\kappa\,\vec{n}$, $d\vec{n}/dx=-\kappa\,\vec{t}+\tau\,\vec{b}$
and $d\vec{b}/dx=-\tau\,\vec{n}$.

We first introduce a curvilinear coordinate system. The position of a point
of space is specified by coordinates $(x,y,z)$ through
\begin{equation}\label{gol1}
\vec{r}(x,y,z) = \vec{r}_{\sss C} (x) + y\, \vec{N} + z \, \vec{B} \; ,
\end{equation}
where $\vec{N}(x)= \cos\theta\,\vec{n} + \sin\theta\,\vec{b}$,
$\vec{B}(x)= -\sin\theta\,\vec{n} + \cos\theta\,\vec{b}$ and $\theta(x)$ is
defined through $d\theta/dx=-\tau(x)$.

We then select a potential of the form $V_\perp (y,z)$. Note that the choice
of vectors $\vec{N}$ and $\vec{B}$ for defining the transverse coordinates $y$
and $z$ is not irrelevant, i.e., the manner in which $V_\perp$ winds round
${\cal C}$ does matter: it has to be the same as the way $(\vec{N},\vec{B})$
winds around $\vec{t}$. Indeed one can see that some torsion may create a
repulsive potential along $x$ which could cancel the localizing effect of the
bend. This is avoided with the type of coordinate dependence we have chosen. A
simple way of seeing this is by noticing that $\vec\nabla$ has coordinates
$(h^{-1}\partial_x,\partial_y,\partial_z)$ in the
$(\vec{t},\vec{N},\vec{B}\,)$ frame (with $h(x,y,z) =
1-\kappa\,(y\,\cos\theta-z\,\cos\theta)$) and the force $-\vec\nabla\,
V_\perp(y,z)$ thus has no tangential component. It would have been more
natural to define the $(y,z)$ coordinates as in (\ref{gol1}) but using the
$(\vec{n},\vec{b}\,)$ vectors instead of $(\vec{N},\vec{B}\,)$. In this case
however the force $-\vec\nabla\, V_\perp(y,z)$ would generically have a
tangential component which could spoil the localizing properties of the bend.
Note that the latter discussion is of course irrelevant for a potential
$V_\perp$ with circular symmetry (as any simple experimental wave guide is
expected to have).

Denoting $h'=\partial_{x}h$ and $h''=\partial^2_{xx}h$, the Schr\"odinger
equation for $\Phi=h^{1/2}\,\Psi$ reads
\begin{equation}\label{gol4}
-\frac{1}{2}\left(
\frac{1}{h^2}\partial^2_{xx} +\partial^2_{yy} +\partial^2_{zz}
\right)\,\Phi
+\frac{h'}{h^3}\, \partial_x\Phi
+\left[
-\frac{\kappa^2}{8\,h^2} - \frac{5\,(h')^2}{8\, h^4}
+\frac{h''}{4\,h^3} + V_\perp (y,z) 
\right] \, \Phi = \mu \, \Phi \; .
\end{equation}
Note that in (\ref{gol4}) as in the rest of the paper, we take
$\hbar=m=1$. The choice of coor\-di\-na\-tes (\ref{gol1}) gives a volume
element $d^3r=h\,dx\,dy\,dz$ and $\Phi$ is thus normalized as
$\int dx\,dy\,dz |\Phi|^2=1$.

The adiabatic limit is defined by $h\to 1$, $\kappa \gg h'$, $\kappa \gg
\sqrt{h''}$. In this limit Eq.~(\ref{gol4}) decouples into a longitudinal and
a transverse equation. One obtains for the longitudinal equation a potential
$V_\parallel(x)=-\kappa^2(x)/8$, attractive in the region of the bend. Since
any attractive potential in one dimension has a bound state \cite{Lan}, in
this limit there exists a quantum state localized in the bend.

The theorem of Goldstone and Jaffe \cite{Gol92} establishes the existence of a
bound state for much more general wave-guides, with arbitrary curvature. It
was originally demonstrated for sharp-walls wave guides, but the proof can be
straightforwardly extended to the case of a smooth confining potential. Hence
we do not reproduce its derivation here. We just note that for some points of
space, the coordinate system ({\ref{gol1}) can be ambiguous. This imposes for
the sharp wall problem the requirement that the transverse size of the guide
does not exceed the radius of curvature of ${\cal C}$. For the smooth
potential problem, the same restriction requires that the typical range of
$V_\perp$ (or equivalently the spatial extension of $\phi_\perp$) is lower
than the radius of curvature of ${\cal C}$. Obviously this condition is much
weaker than the condition of adiabaticity. Note also that the Goldstone-Jaffe
theorem has the same limitation as above, namely some torsion may destroy the
localizing effect of the bend and the way $V_\perp$ winds around the curve
${\cal C}$ is not irrelevant. A potential $V_\perp(y,z)$ where the coordinates
$y$ and $z$ are defined above (Eqs.~(\ref{gol1})) ensures the applicability of
the theorem.


\begin{thebibliography}{99}
\bibitem{foot4} Careful derivation shows that (\ref{bound4}) is applicable for
$N=0$ in the limit $\sigma^{-2} \gg V_0$, where
$V_0$ and $\sigma$ are typical orders of magnitude for
the depth and range of $V_\parallel(x)$, respectively. This condition
corresponds to shallow potentials with a zero point energy much larger than
$V_0$. If this condition is fulfilled for $N=0$, increasing $N$
increases the spatial extension of the bound state and makes the approximation
even better justified. When $N$ tends to $N_{max}$ the approximation is
valid for any type of potential (not necessarily shallow).
\bibitem{refadia} As discussed at the end of Sec.~\ref{sec2},
such bendings have been treated successfully for linear waves
in Ref.~\cite{Spr92}
with an adiabatic approximation of the type (\ref{adia1}).
\end{thebibliography}
\end{document}